\documentclass{amsart} 

\usepackage{amsmath}
\usepackage{hyperref}
\usepackage{gensymb} 

\usepackage{natbib}

\usepackage{graphicx}
\usepackage{longtable}
\usepackage{multirow}
\usepackage{eurosym}
\usepackage{xcolor,colortbl}
\usepackage{amsmath}
\usepackage{float}
\usepackage[caption = false]{subfig}
\definecolor{Gray}{gray}{0.85}
\newcolumntype{a}{>{\columncolor{Gray}}p}
\let\vec=\mathbf
\newcommand{\unit}[1]{\hat{\vec{#1}}}

\begin{document}

		\title{Light radiation pressure upon an optically orthotropic surface}

		\author{Nikolay A. Nerovny}
		\email{nick.nerovny@bmstu.ru}
		
		\author{Irina E. Lapina}
		\email{iealpina8@gmail.com}
		
		\author{Anton S. Grigorjev}
		\email{baldy.ash@yandex.ru}

		\address{Bauman Moscow State Technical University, 5 stroenie 1 2-ya Baumanstaya st., Moscow, 105005, Russia}
		
		\begin{abstract}
			In this paper, we discuss the problem of determination of light radiation pressure force upon an anisotropic surface.
			The optical parameters of such a surface are considered to have major and minor axes, so the model is called an orthotropic model.
			We derive the equations for force components from emission, absorption, and reflection, utilizing a modified Maxwell's specular-diffuse model.
			The proposed model can be used to model a flat solar sail with wrinkles.
			By performing Bayesian analysis for example of a wrinkled surface, we show that there are cases in which an orthotropic model of the optical parameters of a surface may be more accurate than an isotropic model.
			
		\end{abstract}

	\maketitle
	
	\allowdisplaybreaks

	\section*{Nomenclature}
	
	$O'x'_1x'_2x'_3$ -- global coordinate frame
	
	$Ox_1x_2x_3$ -- local coordinate frame
	
	$\unit{e}'_i,\ i=1,2,3$ -- unit vectors of global coordinate frame
	
	$\unit{e}_i,\ i=1,2,3$ -- unit vectors of local coordinate frame
	
	$dA$ -- infinitesimal element of surface $A$
	
	$\unit{n}$ -- normal to $dA$
	
	$\unit{m}$ -- orientation vector for orthotropic model (in plane $Ox_1x_2$)
	
	$\vec{r}$ -- position of $dA$ in global frame
	
	$\theta,\beta$ -- direction angles in local frame
	
	$\epsilon'_\lambda$ -- directional spectral emissivity
	
	$\epsilon'$ -- directional integral emissivity
	
	$\epsilon$ -- emissivity (for isotropic case)
	
	$B$ -- Lambertian coefficient
	
	$T$ -- temperature of $dA$
	
	$\epsilon_1,\ \epsilon_2,\ \theta_m$ -- parameters of orthotropic model for emission
	
	$B_m$ -- modified Lambertian coefficient for orthotropic emission
	
	$c$ -- speed of light in vacuum
	
	$i'^A_\lambda$ -- directional spectral intensity of irradiation
	
	$i'^A$ -- directional integral intensity of irradiation
	
	$q_0$ -- integral intensity of light source
	
	$\unit{s}$ -- vector from light source to $dA$
	
	$\rho''_\lambda$ -- bidirectional spectral reflectivity
	
	$\rho''$ -- bidirectional integral reflectivity
	
	$I'$ -- hemispherical-directional light intensity
	
	$s$ -- specularity coefficient
	
	$\rho$ -- reflectivity (for isotropic model)
	
	$\rho_1,\ \rho_2,\ \theta_m$ -- parameters for orthotropic model for reflection
	
	$B_\rho$ -- modified Lambertian coefficient for orthotropic reflection
	
	$d\mathbf{F}^{Sr}$ -- fraction of emission pressure in arbitrary direction $\unit{r}$
	
	$d\mathbf{F}^{Ar}$ --fraction of absorption pressure in arbitrary direction $\unit{r}$
	
	$d\vec{F}^{Rr}$ -- fraction of reflection pressure in arbitrary direction $\unit{r}^R$
	
	$d\mathbf{F}^{S}$ -- light pressure from emission
	
	$d\mathbf{F}^{A}$ -- light pressure from absorption
	
	$d\vec{F}^{R}$ -- light pressure from reflection
	
	$d\vec{F}$ -- total light radiation pressure upon $dA$
	
	\section*{Introduction}
	
	The theory of light radiation pressure upon space objects is well-established.
	For celestial bodies, this pressure creates the Yarkovsky acceleration due to uneven heating of their surface~\cite{vokrouhlicky_yarkovsky_1998,hartmann_reviewing_1999}.
	There is also a Yarkovsky-O'Keefe–Radzievskii–Paddack (YORP) effect, in which an asteroid can spin-up from emission pressure because of its irregular shape,~\cite{vokrouhlicky_yorp-induced_2002} up to the disintegration of a body~\cite{paddack_rotational_1969,rubincam_radiative_2000}.
	
	For practical applications, the derivation of light radiation pressure force is necessary for the prediction of the dynamics of GNSS satellites~\cite{fliegel_solar_1996,bar-sever_new_1997,springer_new_1999,bar-sever_new_2004,rodriguez-solano_adjustable_2012,tan_new_2016}, for interplanetary stations~\cite{kubo-oka_solar_1999,vaughan_momentum_2001,turyshev_support_2012} and other spacecraft~\cite{kinzel_jwst_2010}.
	
	For solar sail applications, there are many studies of light radiation pressure, including light pressure generalizations~\cite{forward_1989,mcinnes_solar_2004}, and special cases -- variable reflectance / transmittance coatings~\cite{kislov_2004}, degradation effects~\cite{dachwald_potential_2005,dachwald_parametric_2006}, joint analysis of aerodynamic and radiation forces on spacecraft~\cite{shmatov_joint_2014}, laser propulsion~\cite{forward_roundtrip_1984,popova_stability_2016}, transparent sails~\cite{swartzlander_jr_radiation_2017}, etc.
	There are numerous studies of the astrodynamics of solar sails~\cite{farres_dynamics_2016,gachet_geostationary_2016,lachut_towards_2016,ono_generalized_2016,felicetti_attitude_2017,german_satellite_2017,ma_controllable_2017,niccolai_analytical_2017} etc.
	
	In the space experiments Nanosail-D2 \cite{alhorn_nanosail-d_2011}, IKAROS \cite{tsuda_flight_2011,ikaros1}, and LightSail \cite{ridenoure2015lightsail} it was shown that any solar sail membrane has general curvature, both regular and semi-random (smoothness), and also small wrinkles.
	
	The light pressure model on the curved solar sail was generalized by~\cite{rios-reyes_applications_2004,rios_reyes_generalized_2005,rios-reyes_2006,rios-reyes_solar-sail_2007,scheeres_dynamical_2007,mcmahon_new_2010,mcmahon_general_2014,mcmahon_improving_2015} and extended by~\cite{jing_curved_2012,jing_solar_2014,nerovny_representation_2017}.
	This model is called the Generalized Sail Model (GSM).
	
	In this paper, we will consider the optical anisotropy from the geometrical sources of this anisotropy.
	The main sources of this optical anisotropy are wrinkles on the solar sail membrane~\cite{wong_wrinkled_2006,jenkins_recent_2006}.
	One special case of the effects of wrinkles on the solar sail efficiency was studied by~\cite{greschik_direct_2014-1}.
	
	We will derive the equations for light radiation pressure by utilizing the well-established theory of light-matter interaction as in radiative heat transfer \cite{howell_thermal_2015}, and after this, we will move to the vector representation of force.
	
	We will consider the effects of emission, absorption, and reflection on light pressure because further phenomena such as transmission are supposed to be less influential on solar sails than the main effects~\cite{forward_1989}. 
	For each effect, we will consider both isotropic and anisotropic cases.
	For reflection, we will utilize Maxwell's reflection model \cite{howell_thermal_2015}, in which we assume that reflection has two components as a sum of diffuse and specular cases with corresponding specularity coefficient $s$.
	We will also consider the back reflection phenomenon for the orthotropic model.
	
	\section{Reference frames}
	
\begin{figure}
		\centering
		\includegraphics[width = 90mm]{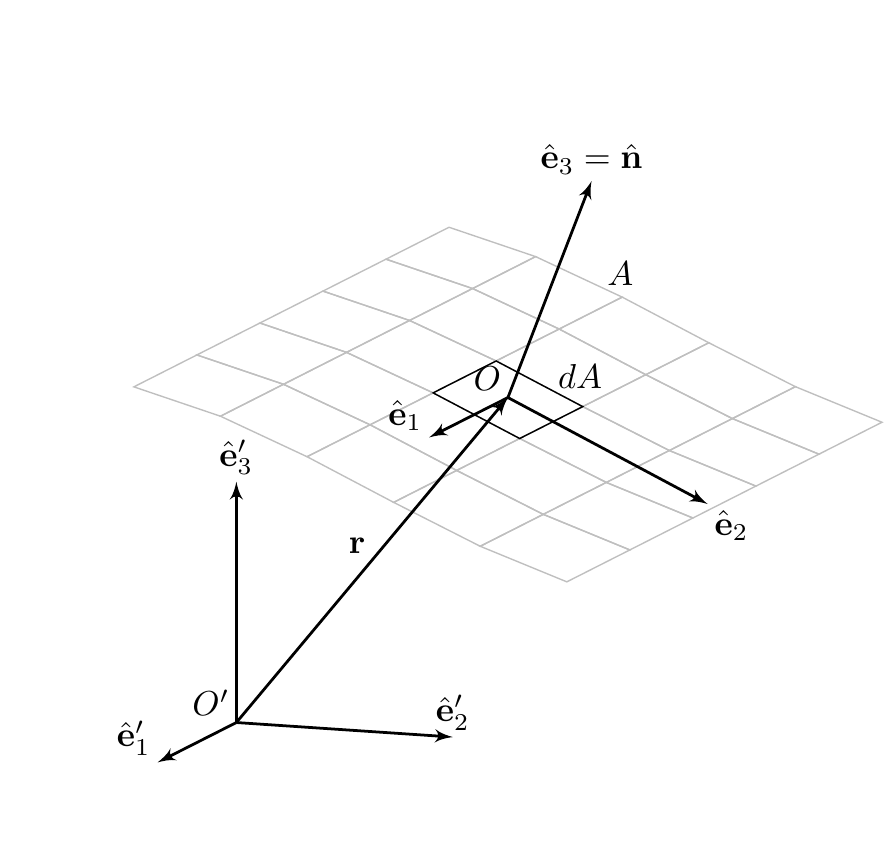}
		\caption{Definition of coordinate frames}\label{f:shape}
	\end{figure}

	Let us consider some surface $A$ in Euclidean space with origin $O'$ and associated Cartesian coordinate system $O'x'_1x'_2x'_3$, Fig.~\ref{f:shape}.
	We will call this frame a global frame.
	Let us introduce $\unit{e}'_i$ -- unit vectors for the global frame, $i = 1{,}\ 2{,}\ 3$.
	
	On this surface, it is possible to localize an infinitesimal surface element $dA$ for which we introduce a local Cartesian coordinate system $Ox_1x_2x_3$ with unit vectors $\unit{e}_i$, Fig.~\ref{f:shape}.
	The origin of local frame $O$ is situated in the center of $dA$, and its normal $\unit{n}$ is equal to $\unit{e}_3$.
	$[T]$ is a transformation matrix from the local frame to the global frame.
	The orientation of $Ox_1$ and $Ox_2$ is arbitrary.
	
\begin{figure}
		\centering
		\includegraphics[width = 60mm]{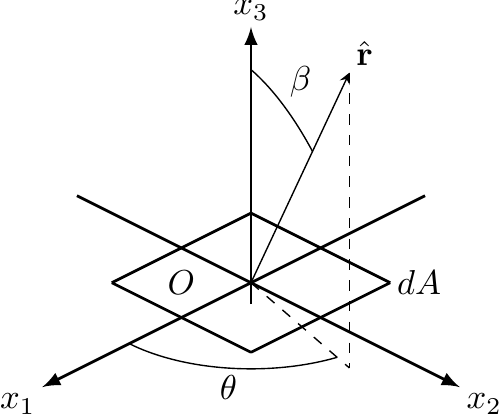}
		\caption{Definition of angles for arbitrary unit vector $\unit{r}$}\label{f:define_1}
	\end{figure}

	In the following equations, for any vector, e.g. $\unit{r}$, we will use direction angles $(\beta,\theta)$ in the local frame as follows (Fig.~\ref{f:define_1}):
	\begin{itemize}
		\item $\beta\in[0,\pi/2]$ -- angle between vector $\unit{r}$ and $+x_3$. We consider that the infinitesimal surface element is laying on the plane $Ox_1x_2$.
		\item $\theta\in[0,2\pi]$ -- angle between axis $Ox_1$ and a projection of $\unit{r}$ on the plane $Ox_1x_2$, counterclockwise around $Ox_3$.
	\end{itemize}
	
	Direction angles $(\beta,\theta)$ may have additional subscripts or superscripts.
	
	%
	%
	
	\section{Model}
	
	\subsection{Thermal emission}
	
	Let $\epsilon'_\lambda$ be a directional spectral emissivity, which depends on wavelength, temperature, and shows the difference of emission of $dA$ in direction $(\beta,\theta)$ as compared with black body emission in the same direction.  
	One can write the equation of directional integral emissivity~\cite{howell_thermal_2015}:
	\[
	\label{eq:napr_eps}
	\epsilon'(\beta,\theta,T) = \frac{\pi \int\limits_0^\infty {\epsilon'_\lambda i'_{\lambda b}}d\lambda}{\sigma T^4},
	\]
	Where $\sigma$ -- Stephan-Boltzmann constant, $i'_{\lambda b}(\lambda,T)$ -- spectral intensity of blackbody radiation which is represented by Planck's law:
	\[
	\label{eq:planck}
	i'_{\lambda b}(\lambda,T) = \frac{2hc^2}{\lambda^5 {\left( e^{\frac{hc}{\lambda k T}} -1 \right)}},
	\]
	Where $h$ -- Planck's constant, $c$ -- light speed in vacuum, $k$ -- Boltzmann constant.
	
	Let us introduce an arbitrary unit vector $\unit{r}$ in the local frame:
	\[
	\unit{r}(\beta,\theta) = \sin{\beta}\cos{\theta}\unit{e}_1 + \sin{\beta}\sin{\theta}\unit{e}_2 + \cos{\beta}\unit{e}_3,
	\]
	Where $\beta\in[0;\pi/2]$ and $\theta\in[0;2\pi]$.

	One can write the equation of the fraction of light radiation pressure in direction $\unit{r}$:
	\[
	d\mathbf{F}^{Sr}(\beta,\theta,T) = -\frac{\epsilon' \sigma T^4}{c}\unit{r} \cos\beta dA.
	\]
	The superscript $S$ stands for emission (``self'').
	The relation $\cos\beta dA$ is the area of the infinitesimal element $dA$ under the angle $\beta$.
	
	The equation for projection of light pressure force from emission in the direction $\unit{e}_i$ on the area $dA$ can be written as follows:
	\begin{equation}
	\label{eq:dF_S1}
	dF^{S}_i(T) = \frac{1}{2\pi}\int\limits_0^{2\pi} \int\limits_0^{\frac{\pi}{2}} d\mathbf{F}^{Sr} \cdot \unit{e}_i  d\beta d\theta dA.
	\end{equation}
	
	\paragraph{Isotropic case}
	
	Let us assume that the optical parameters are independent of direction within the surface i.e. they are isotropic.
	Function $\epsilon'$ can be considered as an axis-symmetrical function around $Ox_3$,
	\begin{equation}
	\epsilon' = \epsilon f_\epsilon(\beta),\label{eq:eps_f}
	\end{equation}
	Where $\epsilon=\text{const}$ -- emissivity of the material.
	
	After rewriting of~\eqref{eq:dF_S1} in corresponding projections, using~\eqref{eq:eps_f}, we can obtain the following relations:
	\begin{align*}
	&dF^S_1 = - \frac{\sigma T^4}{2\pi c}\epsilon\int\limits_0^{\frac{\pi}{2}}f_\epsilon(\beta) \sin\beta\cos\beta\left( \int\limits_0^{2\pi} \cos\theta d\theta \right)d\beta dA = 0;\\
	&dF^S_2 = - \frac{\sigma T^4}{2\pi c}\epsilon\int\limits_0^{\frac{\pi}{2}}f_\epsilon(\beta) \sin\beta\cos\beta\left( \int\limits_0^{2\pi} \sin\theta d\theta \right)d\beta dA = 0;\\
	&dF^S_3 = - \frac{\sigma T^4}{2\pi c}\epsilon\int\limits_0^{\frac{\pi}{2}}f_\epsilon(\beta) \cos^2\beta\left( \int\limits_0^{2\pi} d\theta \right)d\beta dA = -\frac{\sigma T^4}{c}\epsilon\int\limits_0^{\frac{\pi}{2}}f_\epsilon(\beta) \cos^2\beta d\beta dA.\\
	\end{align*}
	
	Now we introduce the coefficient $B$:
	\[
	B = \int\limits_0^{\frac{\pi}{2}}f_\epsilon(\beta) \cos^2\beta d\beta.\label{eq:B_definition}
	\]
	
	This value represents the composition of the axially symmetric radiation pattern of emission.
	It is often called a Lambertian coefficient~\cite{jing_curved_2012}, since its most common value $B=2/3$ corresponds to a Lambertian diffuse surface.
	For diffuse surface $f_\epsilon(\beta)=\cos\beta$ (Lambert's law),
	$$
	B=\int\limits_0^{\frac{\pi}{2}}\cos^3\beta d\beta = \frac{2}{3}.
	$$
	
	We can rewrite the equation for light pressure force from emission using vector notation:
	\begin{equation}
	d\mathbf{F}^S = - \frac{\epsilon B \sigma T^4}{c}\unit{n}.\label{eq:dFS_final}
	\end{equation}
	
	Eq.~\eqref{eq:dFS_final} is independent of the coordinate frame choice, since we can always represent $\unit{n}$ in different coordinate systems using the corresponding transformation matrix~$[T]$.
	
	\paragraph{Orthotropic case}
	
\begin{figure}
		\centering
		\includegraphics[width = 90mm]{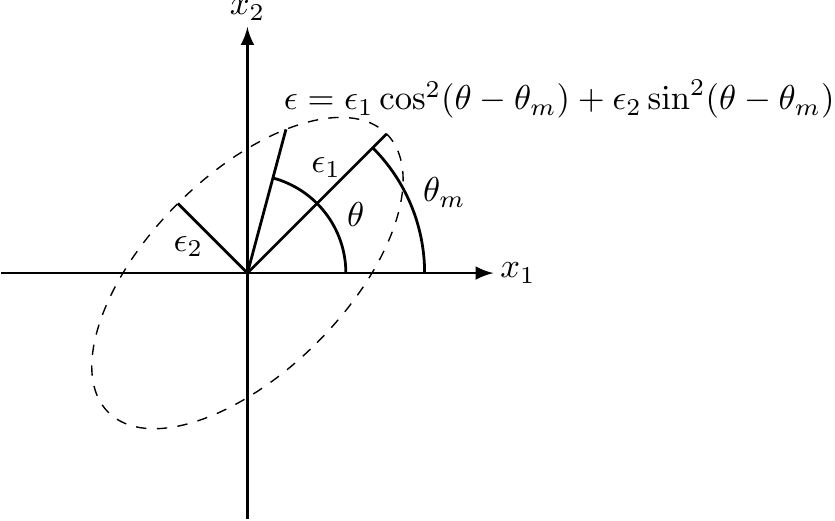}
		\caption{Orthotropic model for emissivity}\label{f:ellipse_eps}
	\end{figure}

	We will utilize the following model of optical parameters.
	There are two axes of optical parameters.
	In the direction that is defined by angle $\theta_m$ in the plane $Ox_1x_2$, the emissivity is equal to $\epsilon_1$.
	In the perpendicular direction, the emissivity is equal to $\epsilon_2$.
	We will assume that emissivity changes analytically according to some law (Fig.~\ref{f:ellipse_eps}).
	So we can write the equation for directional integral emissivity as follows:
	\[
	\epsilon'_m(\beta,\theta) = (\epsilon_1\cos^2(\theta-\theta_m) + \epsilon_2\sin^2(\theta - \theta_m))\cos\beta.
	\]
	
	For the isotropic case, assume $\epsilon_1 = \epsilon_2 = \epsilon$ for any $\theta_m$ $\epsilon'_m(\beta,\theta) = \epsilon\cos\beta$.
	
	Now we can write the projections of light pressure force from heat emission:
	\begin{align*}
	&dF^{Sm}_1 = -\frac{\sigma T^4}{2\pi c} \int\limits_{0}^{2\pi}\int\limits_{0}^{\frac{\pi}{2}}\epsilon'_m(\beta,\theta)\cos\beta \sin\beta \cos\theta d\beta d\theta dA = 0;\\
	&dF^{Sm}_2 = -\frac{\sigma T^4}{2\pi c} \int\limits_{0}^{2\pi}\int\limits_{0}^{\frac{\pi}{2}}\epsilon'_m(\beta,\theta)\cos\beta \sin\beta \sin\theta d\beta d\theta dA = 0;\\
	&dF^{Sm}_3 = -\frac{\sigma T^4}{2\pi c} \int\limits_{0}^{2\pi}\int\limits_{0}^{\frac{\pi}{2}}\epsilon'_m(\beta,\theta)\cos^2\beta d\beta d\theta dA = -\frac{\sigma T^4 (\epsilon_1 + \epsilon_2)}{3 c}dA.
	\end{align*}
	
	Now we can introduce the modified Lambertian coefficient $B_m$ as follows:
	\[
	B_m = \frac{\epsilon_1 + \epsilon_2}{3\epsilon_1}.
	\]
	
	By introducing of the normal unit vector $\unit{n}$, we can write the equation for emission pressure from an orthotropic surface:
	\begin{equation}
	d\mathbf{F}^S = - \frac{\epsilon_1 B_m \sigma T^4}{c}\unit{n}\label{eq:dFS_final_m}
	\end{equation}
	
	for isotropic case, where $\epsilon_1 = \epsilon_2 = \epsilon$, $B_m = 2/3.$
	
	\subsection{Absorption}
	
	For light pressure from absorption, we will consider that all of the light flux linear momenta that fall onto $dA$ are transferred to that surface~\cite{forward_1989} so there is no difference between isotropic and anisotropic models.
	The anisotropy of absorption, however, should be considered in the case of calculation of solar sail temperature.
	
	We can introduce the directional spectral intensity of irradiation $i'^A_\lambda(\lambda,\beta,\theta)$, which depends on the properties and spatial position of the light source relative to $dA$.
	We will also utilize the directional integral intensity of irradiation:
	\begin{equation}
	\label{eq:intens_pad_napr}
	i'^A(\beta,\theta) = \int\limits_0^\infty i'^A_\lambda d\lambda.
	\end{equation}
	
	Light pressure force fraction in direction $\unit{r}$ can be written as follows:
	\begin{equation}
	d\mathbf{F}^{Ar}(\beta,\theta) = -\frac{i'^A}{c}\unit{r} \cos\beta dA.\label{eq:dF_A1}
	\end{equation}
	
	The superscript $A$ stands for the absorption.	
	
	One can write the equation for projection of light pressure force from absorbed radiation in the direction $\unit{e}_i$:
	\begin{equation}
	\label{eq:dF_A}
	dF^A_i = \frac{1}{2\pi} \int\limits_0^{2\pi} \int\limits_0^{\frac{\pi}{2}} d\mathbf{F}^{Ar} \cdot \unit{e}_i   d\beta d\theta.
	\end{equation}
	After substitution of Eq.~\eqref{eq:dF_A1} into Eq.~\eqref{eq:dF_A}, using Eq.~\eqref{eq:intens_pad_napr}, we can get:
	\begin{equation}
	\label{eq:dF_AA}
	dF^A_i  = -hc \int\limits_0^\infty \int\limits_0^{2\pi} \int\limits_0^{\frac{\pi}{2}} \frac{\unit{r} \cdot \unit{e}_i i'^A_\lambda}{2\pi hc^2} \cos\beta d\beta d\theta d\lambda dA.
	\end{equation}
	
	We will only consider the point light source with position $(\beta_0,\theta_0)$.
	The intensity of the falling light can be written as follows:
	\begin{equation}
	\label{eq:sample1}
	i'^A_\lambda = 2\pi q(\lambda)\cos{\beta}\delta(\beta-\beta_0)\delta(\theta-\theta_0),
	\end{equation}
	Where $\delta(x)$ -- Dirac's delta function, $q(\lambda)$ -- spectral intensity of light source for which the integral intensity is equal to $q_0$:
	\begin{equation}
	\label{eq:sample2}
	q_0 = \int\limits_0^\infty q(\lambda) d\lambda.
	\end{equation}
	
	The derivation for light pressure force from non-point light source can be found in~\cite{mcinnes_dynamics_1990,mcinnes_solar_1990}.
	
	After substitution of \eqref{eq:sample1} and \eqref{eq:sample2} into \eqref{eq:dF_AA}, we can get the projections of light pressure force for fully absorbed light:
	\begin{align*}
	&dF^A_1 = -\frac{q_0}{c}\int\limits_0^{\frac{\pi}{2}}\int\limits_0^{2\pi}\sin\beta\cos\beta\cos\theta\delta(\beta-\beta_0)\delta(\theta-\theta_0)d\theta d\beta dA =\\
	&= -\frac{q_0}{c}\sin\beta_0\cos\beta_0\cos\theta_0 dA;\\
	&dF^A_2 = -\frac{q_0}{c}\int\limits_0^{\frac{\pi}{2}}\int\limits_0^{2\pi}\sin\beta\cos\beta\sin\theta\delta(\beta-\beta_0)\delta(\theta-\theta_0)d\theta d\beta dA =\\
	&= -\frac{q_0}{c}\sin\beta_0\cos\beta_0\sin\theta_0 dA;\\
	&dF^A_3 = -\frac{q_0}{c}\int\limits_0^{\frac{\pi}{2}}\int\limits_0^{2\pi}\cos^2\beta\delta(\beta-\beta_0)\delta(\theta-\theta_0)d\theta d\beta dA = -\frac{q_0}{c}\cos^2\beta_0 dA.
	\end{align*}
	
	We can introduce the additional unit vector of light orientation $\unit{s}$, which is pointing from light source to surface element $dA$:
	\[
	\unit{s}(\beta_0,\theta_0) = -\sin{\beta_0}\cos{\theta_0}\unit{e}_1 - \sin{\beta_0}\sin{\theta_0}\unit{e}_2 - \cos{\beta_0}\unit{e}_3,
	\]
	
	After this we can rewrite the equation for light pressure from absorbed light in simplified vector notation:
	\begin{equation}
	d\vec{F}^A = -\frac{q_0}{c}(\unit{n}\cdot\unit{s})\unit{s} dA.\label{eq:dF_AF}
	\end{equation}

	\subsection{Reflection}
	
	Now we will consider the light radiation pressure from reflected light.
	Let us introduce the bidirectional spectral reflectivity $\rho''_\lambda(\lambda,\beta^R,\theta^R,\beta,\theta)$,
	where $(\beta^R,\theta^R)$ is an orientation of reflected intensity.
	Considering the directional spectral intensity $i'^A_\lambda(\lambda,\beta,\theta)$, one can write the bidirectional integral reflectivity:
	\[
	\label{eq:dvunapr}
	\rho''(\beta^R,\theta^R,\beta,\theta) = \frac{\int\limits_0^\infty \rho''_\lambda i'^A_\lambda d\lambda}{i'^A},
	\]
	Where $i'^A$ can be calculated using \eqref{eq:intens_pad_napr}.
	
	Let us introduce the arbitrary unit vector $\unit{r}^R$, which represents the direction of reflected light flux:
	\[
	\unit{r}^R(\beta,\theta) = \sin{\beta^R}\cos{\theta^R}\unit{e}_1 + \sin{\beta^R}\sin{\theta^R}\unit{e}_2 + \cos{\beta^R}\unit{e}_3.
	\]
	
	We can calculate the hemispherical-directional light intensity:
	\[
	I'=\frac{1}{2\pi}\int\limits_0^{\frac{\pi}{2}}\int\limits_0^{2\pi}\rho''i'^A d\theta d\beta.
	\]
	
	The fraction of light pressure force from reflected light in direction $\unit{r}^R$ will be as follows:
	\[
	d\vec{F}^{Rr}=-\frac{I'\unit{r}^R}{c}\cos\beta^R dA,
	\]
	and the total infinitesimal force from reflected light can be calculated for diffuse reflection as the follows:
	\[
	dF^R_i=-\frac{1}{2\pi c}\int\limits_0^{\frac{\pi}{2}}\int\limits_0^{2\pi}I'\unit{r}^R\cdot\unit{e}_i\cos\beta^R d\theta^R d\beta^R dA,
	\]
	and for specular reflection,
	\[
	dF^R_i=-\frac{1}{2\pi c}\int\limits_0^{\frac{\pi}{2}}\int\limits_0^{2\pi}I'\unit{r}^R\cdot\unit{e}_i d\theta^R d\beta^R dA,
	\]
	since all amount of specularly reflected flux will reflect in the same direction, as opposed to the diffuse case, in which we should take into account the projection of $dA$ in the direction perpendicular to $\unit{r}^R$, i.e. $\cos\beta^R dA$.

	\subsubsection{Diffuse reflection}
	\paragraph{Isotropic case}
	
	For axially symmetric (diffuse) reflection the bidirectional reflectivity can be represented by the following equation:
	\[
	\rho''_1(\beta,\theta,\beta^R,\theta^R)=(1-s)\rho f_\epsilon(\beta^R).
	\]
	
	By use of Eq.~\eqref{eq:sample1} we can write the hemispherical-directional intensity:
	\[
	I'_1=(1-s)\rho q_0 f_\epsilon(\beta^R)\cos\beta_0.
	\]
	
	For infinitesimal force we can get the following relation:
	\[
	dF^{R1}_i = -\frac{(1-s)\rho q_0}{2\pi c} \int\limits_0^{\frac{\pi}{2}}\int\limits_0^{2\pi} f_\epsilon(\beta^R)\unit{r}^R\cdot\unit{e}_i\cos\beta^R d\theta^R d\beta^R dA,
	\]
	moreover, in coordinate projections we can get:
	\begin{align*}
	&dF^{R1}_1 = 0;\\
	&dF^{R1}_2 = 0;\\
	&dF^{R1}_3 = -\frac{q_0}{c}\rho(1-s)B\cos\beta_0 dA.
	\end{align*}
	
	By utilizing vector notation, after introducing the light source orientation vector $\unit{s}$, we can obtain the equation of light pressure force for diffusely reflected light:
	\begin{equation}
	d\vec{F}^{R1}=\frac{q_0}{c}\rho(1-s)B(\unit{n}\cdot\unit{s})\unit{n}.\label{eq:dF_R1}
	\end{equation}
	
	\paragraph{Orthotropic case}
	
\begin{figure}
		\centering
		\includegraphics[width = 90mm]{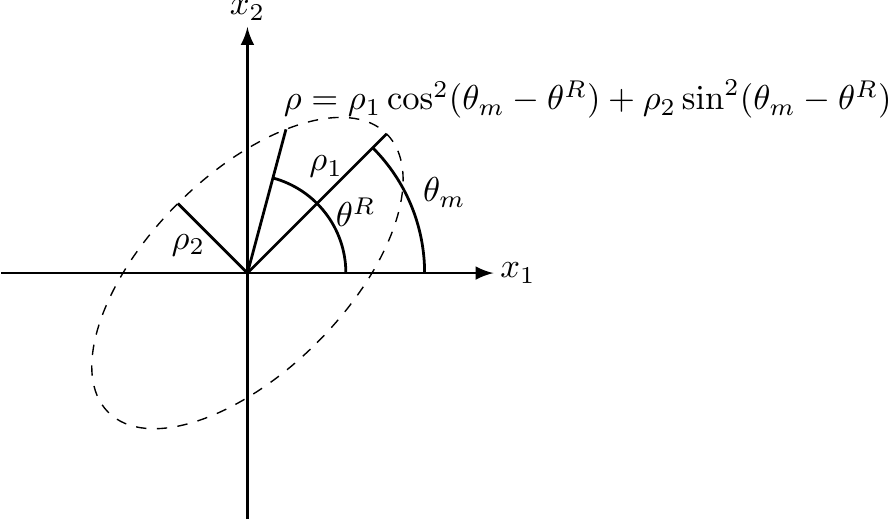}
		\caption{Orthotropic model for reflectivity}\label{f:ellipse_rho}
	\end{figure}

	For the orthotropic case the bidirectional reflectivity will be modeled as the following relation (Fig.~\ref{f:ellipse_rho}):
	\[
	\rho''_{1m}(\beta,\theta,\beta^R,\theta^R)=2\pi(1-s) (\rho_1\cos^2(\theta_m-\theta^R) + \rho_2\sin^2(\theta_m - \theta^R)) \cos\beta^R.
	\]
	
	For specularity $s$ we will also utilize the similar orthotropic model:
	\[
	s = s_1\cos^2(\theta_m-\theta^R) + s_2\sin^2(\theta_m - \theta^R).
	\]
	
	The hemispherical-directional intensity can be obtained by integration:
	\begin{align*}
	&I'_1=\frac{1}{2\pi} \int\limits_{0}^{2\pi}\int\limits_{0}^{\frac{\pi}{2}}\int\limits_{0}^{\infty}\rho''_{1m} i'^A_\lambda d\lambda d\beta d\theta =\\
	&= q_0(1-s_1\cos^2(\theta_m-\theta^R) - s_2\sin^2(\theta_m - \theta^R))\\
	&(\rho_1\cos^2(\theta_m-\theta^R) + \rho_2\sin^2(\theta_m - \theta^R))\cos\beta_0\cos\beta^R.
	\end{align*}
	
	After integration, we can get the following projections of force:
	\begin{align*}
	&dF^{R1m}_1 = 0;\\
	&dF^{R1m}_2 = 0;\\
	&dF^{R1m}_3 = -\frac{q_0}{c}(1-s_1)\rho_1 B_\rho\cos\beta_0 dA,
	\end{align*}
	Where
	\[
	B_\rho = \frac{(4-3s_1-s_2)\rho_1 + (4-s_1-3s_2)\rho_2}{12(1-s_1)\rho_1}.
	\]
	
	For the isotropic case $\rho_1 = \rho_2 = \rho$, $s_1=s_2=s$, $B_\rho = B = 2/3$.
	
	The equation for light pressure force from diffusely reflected light from an orthotropic surface can be represented in the following vector form:
	\begin{equation}
	d\vec{F}^{R1}=\frac{q_0}{c}\rho(1-s)B_\rho(\unit{n}\cdot\unit{s})\unit{n}.\label{eq:dF_R1_m}
	\end{equation}

	\subsubsection{Specular reflection}
	\paragraph{Isotropic case}
	
	For the specular case we can introduce the bidirectional reflectivity as follows:
	\[
	\rho''_2 = 2\pi\rho s \delta(\beta^R-\beta_0)\delta(\theta^R-\theta_0-\pi).
	\]
	
	After integrating we can obtain the hemispherical-directional reflectivity:
	\[
	I'_2 = 2\pi q_0 \rho s \delta(\beta^R-\beta_0)\delta(\theta^R - \theta_0 - \pi)\cos\beta_0.
	\]
	
	We can simply evaluate the following equations for projections of light pressure force:
	\begin{align*}
	&dF^{R2}_1 = \frac{q_0\rho s}{c}\cos\theta_0\sin\beta_0\cos\beta_0 dA;\\
	&dF^{R2}_2 = \frac{q_0\rho s}{c}\sin\theta_0\sin\beta_0\cos\beta_0 dA;\\
	&dF^{R2}_3 = -\frac{q_0\rho s}{c}\cos^2\beta_0 dA,
	\end{align*}
	in vector notation
	\begin{equation}
	d\vec{F}^{R2} = \frac{q_0\rho s}{c}\left( (\unit{n}\cdot\unit{s})\unit{s} - 2(\unit{n}\cdot\unit{s})^2\unit{n} \right)dA.\label{eq:dF_R2}
	\end{equation}

	\paragraph{Orthotropic case}
	
	For the orthotropic case the bidirectional reflectivity will be as follows:
	\begin{align*}
	&\rho''_2 =  2\pi \delta(\beta^R-\beta_0)\delta(\theta^R-\theta_0-\pi)(\rho_1\cos^2(\theta_m-\theta^R) + \rho_2\sin^2(\theta_m - \theta^R))\\
	&(s_1\cos^2(\theta_m-\theta^R) + s_2\sin^2(\theta_m - \theta^R)).
	\end{align*}
	
	The projections of light pressure from specularly reflected light will be:
	\begin{align*}
	&dF^{R2m}_1=\frac{q_0}{c}(\rho_1\cos^2(\theta_m-\theta_0) + \rho_2\sin^2(\theta_m - \theta_0)) (s_1\cos^2(\theta_m-\theta_0) + s_2\sin^2(\theta_m - \theta_0))\\
	&\cos\theta_0\sin\beta_0\cos\beta_0 dA;\\
	&dF^{R2m}_2=\frac{q_0}{c}(\rho_1\cos^2(\theta_m-\theta_0) + \rho_2\sin^2(\theta_m - \theta_0)) (s_1\cos^2(\theta_m-\theta_0) + s_2\sin^2(\theta_m - \theta_0))\\
	&\sin\theta_0\sin\beta_0\cos\beta_0 dA;\\
	&dF^{R2m}_3=-\frac{q_0}{c}(\rho_1\cos^2(\theta_m-\theta_0) + \rho_2\sin^2(\theta_m - \theta_0)) (s_1\cos^2(\theta_m-\theta_0) + s_2\sin^2(\theta_m - \theta_0))\\
	&\cos^2\beta_0 dA.
	\end{align*}
	
	We can introduce the unit vector of orientation of reflection axes in the local frame:
	\[
	\unit{m}=\cos\theta_m\unit{e}_1 + \sin\theta_m\unit{e}_2.
	\]
	
	After transformations, we can write the following equation for infinitesimal light pressure force from specularly reflected light from an optically orthotropic surface:
	\begin{align}
	&d\vec{F}^{R2m} = \frac{q_0}{c}\frac{(\unit{n}\cdot\unit{s})\unit{s} - 2(\unit{n}\cdot\unit{s})^2\unit{n}}{(1-(\unit{n}\cdot\unit{s})^2)^2}(\rho_1(\unit{m}\cdot\unit{s})^2 + \rho_2((\unit{m}\times\unit{s})\cdot\unit{n})^2)\nonumber\\
	&(s_1(\unit{m}\cdot\unit{s})^2 + s_2((\unit{m}\times\unit{s})\cdot\unit{n})^2).\label{eq:dF_R2_m}
	\end{align}
	
	For an orthotropic case of specular reflection, we also introduce the term from back reflection.
	The bidirectional reflectivity will be:
	\begin{align*}
	&\rho''_{2b} =  2\pi \delta(\beta^R-\beta_0)\delta(\theta^R-\theta_0-\pi)(\rho_1\cos^2(\theta_m-\theta^R) + \rho_2\sin^2(\theta_m - \theta^R))\\
	&\frac{(s_1\cos^2(\theta_m-\theta^R) + s_2\sin^2(\theta_m - \theta^R))}{\cos\beta^R}k,
	\end{align*}
	Where $k$ -- some empirical parameter.
	
	The projections of light pressure from back reflected light will be:
	\begin{align*}
	&dF^{R2b}_1=-\frac{q_0}{c}(\rho_1\cos^2(\theta_m-\theta_0) + \rho_2\sin^2(\theta_m - \theta_0)) (s_1\cos^2(\theta_m-\theta_0) + s_2\sin^2(\theta_m - \theta_0))\\
	&\cos\theta_0\sin\beta_0 k dA;\\
	&dF^{R2b}_2=-\frac{q_0}{c}(\rho_1\cos^2(\theta_m-\theta_0) + \rho_2\sin^2(\theta_m - \theta_0)) (s_1\cos^2(\theta_m-\theta_0) + s_2\sin^2(\theta_m - \theta_0))\\
	&\sin\theta_0\sin\beta_0 k dA;\\
	&dF^{R2b}_3=-\frac{q_0}{c}(\rho_1\cos^2(\theta_m-\theta_0) + \rho_2\sin^2(\theta_m - \theta_0)) (s_1\cos^2(\theta_m-\theta_0) + s_2\sin^2(\theta_m - \theta_0))\\
	&\cos\beta_0 k dA.
	\end{align*}
	
	In vector notation:
	\begin{align}
	&d\vec{F}^{R2b} = \frac{q_0}{c}\frac{k \unit{s}}{(1-(\unit{n}\cdot\unit{s})^2)^2}(\rho_1(\unit{m}\cdot\unit{s})^2 + \rho_2((\unit{m}\times\unit{s})\cdot\unit{n})^2)\nonumber\\
	&(s_1(\unit{m}\cdot\unit{s})^2 + s_2((\unit{m}\times\unit{s})\cdot\unit{n})^2).\label{eq:dF_R2_mback}
	\end{align}
	
	We included the back-reflection term according to numerical simulations of some wrinkled surface.
	The parameter $k$ of back reflection is a model constant, and it is not necessary that it have a strong physical background.
	In the Discussion section, we will obtain the value of $k$ simultaneously with the other optical parameters.
	We should note that for the isotropic model, $k$ should always be equal to 0.
	The absence of proper physical background for parameter $k$ is a major disadvantage of proposed model. However, it provides better coincidence with ray tracing results. See Discussion section for comparison between optically orthotropic model and ray tracing simulation for some wrinkled surface.
	
	\subsection{Equation of infinitesimal force and torque}
	\paragraph{Isotropic case}
	
	By substitution of terms from emission~\eqref{eq:dFS_final}, absorption~\eqref{eq:dF_AF}, diffuse~\eqref{eq:dF_R1} and specular reflection~\eqref{eq:dF_R2}, we can obtain the total light pressure infinitesimal force of light pressure on element $dA$ for isotropic case:
	\begin{equation}
	d\vec{F}=\frac{q_0}{c}\left[ -\frac{\epsilon B\sigma T^4}{q_0}\unit{n}-(1-\rho s)(\unit{n}\cdot\unit{s})\unit{s} + \rho(1-s)B(\unit{n}\cdot\unit{s})\unit{n} -2\rho s(\unit{n}\cdot\unit{s})^2\unit{n} \right]dA,\label{eq:dF_full_first}
	\end{equation}
	and, by transition from the local frame to the global frame using transformation matrix $[T]$, we can obtain the same equation for infinitesimal force in the global frame.
	
	The obtained equation is similar to the solar sail light pressure force equation that is widely used~\cite{forward_1989,mcinnes_solar_2004,rios_reyes_generalized_2005,jing_solar_2014}; however we consider the emission only from the front surface.
	It is possible to introduce the emission pressure from the back side, i.e.
	\[
	d\mathbf{F}^S_\text{back}=\frac{\epsilon_\text{back}B_\text{back}\sigma T^4}{c}\unit{n}.
	\]
	However, it is not always necessary, because this can be a formulation of light pressure not only upon the solar sail but some structure with large internal volume (i.e. not a thin film). For solar sails, there is another approach, in which one can consider two opposite sides of the solar sail as two different surfaces very close to each other, introducing the visibility function (whether a particular side is illuminated or not).
	In this approach, it is not necessary to explicitly define which side is front and which is back, but the equation for light pressure force has much-complicated form \cite{nerovny_representation_2017}.
	
	It is also possible to obtain the temperature $T$ of $dA$ using the thermal flux equilibrium equation and write the emission term without using $T$, e.g.~\cite{rios-reyes_2006}.
	
	\paragraph{Orthotropic case}
	
	The equation for infinitesimal light pressure force can be summarized using equations \eqref{eq:dFS_final_m}, \eqref{eq:dF_AF}, \eqref{eq:dF_R1_m}, \eqref{eq:dF_R2_m} and~\eqref{eq:dF_R2_mback}.
	
	\paragraph{Light pressure torque}
	
	In both cases, isotropic and orthotropic, the infinitesimal light pressure torque can be obtained by the following cross product:
	\[
	d\vec{M} = \vec{r}\times d\vec{F},
	\]
	Where $\vec{r}=(r'_1,r'_2,r'_3)^T$ -- vector from the origin of the global frame $O'x'_1x'_2x'_3$ to the infinitesimal area $dA$.
	
	\section{Discussion}
	
	\subsection{Bayesian analysis of parameters of the models}  
	
	\paragraph{Bayesian analysis of parameters}
	
	In this section, we will utilize the so-called Bayesian analysis of model parameters~\cite{kruschke_doing_2015,kruschke_bayesian_2017}.
	In this method, one has to follow several steps to determine which model is more probable than another and which parameters of models are more probable.
	These are the steps of Bayesian analysis~\cite{kruschke_doing_2015}:
	\begin{enumerate}
		\item Specify the experimental data, its scales, and identify which values should be predicted by the model and which data are predictors.
		\item Define the descriptive analytical models of the data.
		\item Specify the prior distribution for parameters.
		\item Specify a likelihood function and use Bayesian inference to calculate the posterior probability distribution of parameter values.
		\item Check that the posterior distribution approximates the data.
	\end{enumerate}
	
	\paragraph{Data}
	For this analysis, we obtain the probability distribution of optical parameters of models for optically isotropic ($\rho,\ s,\ B$) and orthotropic surfaces ($\rho_1,\ \rho_2,\ s$).
	We do not consider light pressure force from emission.
	We do not divide values of force by speed of light $c$ and multiply them by flux $q_0$; thus, dimensions of force in all following tables and figures are $m^2$.
	To compare these models, we introduce the model index $m$.
	For the likelihood function, we also introduce the standard deviation $\sigma_F$.
	
	\paragraph{Ray tracing}
	
	We wrote software for direct Monte Carlo simulation of ray tracing.
	This software represents the surface as a mesh of triangles.
	We utilize Maxwell's model of specular-diffuse reflection.
	The diffuse reflection is Lambertian.
	This software does not consider spectral parameters of a light source.
	Calculation of light pressure force was done similar to works~\cite{nerovny_representation_2017,ziebart_generalized_2004}.
	We tested this software on several simple geometries which have analytical models of light radiation pressure: specular solar sail, Lambertian diffuse solar sail, specular and diffuse sphere, cylinder and cone.
	This software is freely available on GitHub~\cite{nerovny_locutus3009/srp2:_2017} under GPLv3 license.
	
	\paragraph{Models}
	We use two models: a model of an optically isotropic surface and a model of an optically orthotropic surface, $\unit{m}=(1,0,0)^T$ in local frame.
	Both models should predict vectors of light radiation pressure $\vec{F}_{\text{I}}$ and $\vec{F}_{\text{O}}$ consequently, for different orientations of a light source, and these vectors should be as close as possible to data $\vec{F}$, which is obtained by direct ray tracing.

	\paragraph{Prior distribution}
	
	Prior distribution for the isotropic model: $\rho,\ s,\ B$ is uniform from 0 to 1, $\sigma_F$ is exponential with parameter equal to 1~\cite{plummer_jags_2013}.
	Prior distribution for the orthotropic model: $\rho_1,\ \rho_2,\ s_1,\ s_2$ is uniform from 0 to 1, $\sigma_F$ and $k$ are exponential with parameter equal to 1.
	Prior distribution for model index $m$: generally, for the isotropic model $P(m=1)=0.5$, for the orthotropic model $P(m=2)=0.5$.
	In subsequent analysis, we use different values for prior probabilities of a model index.
	
	\paragraph{Likelihood function}
	
	The likelihood probability distribution function for this model is the following formula:  
	\[
	P(D|m,\sigma_F,\dots)=\frac{1}{\sqrt{2 \pi} \sigma_F C_c}\left\{\begin{array}{l l}
	e^{-\frac{|\vec{F} - \vec{F}_{\text{I}}|^2}{2\sigma_F^2}}, & m=1;\\
	e^{-\frac{|\vec{F} - \vec{F}_{\text{O}}|^2}{2\sigma_F^2}}, & m=2,
	\end{array} \right.
	\]
	Where: $C_c$ -- an integer large enough to ensure that the likelihood is less than 1 (required by JAGS~\cite{plummer_jags_2013}); $\vec{F}$ -- data from direct Monte Carlo simulation (resultant light pressure force); $\vec{F}_{\text{I}}$ -- predicted vector of light pressure force for given set of parameters considering the isotropic case, Eq.~\eqref{eq:dF_full_first}; $\vec{F}_{\text{O}}$ -- predicted vector of light pressure force for given set of parameters considering orthotropic case, Eq.~\eqref{eq:dF_AF}, \eqref{eq:dF_R1_m} and~\eqref{eq:dF_R2_m}; $\sigma_F$ -- standard deviation.
	
	\paragraph{Posterior distribution}
	
	In our analysis, we use the R code combined with the JAGS toolkit for Markov Chain Monte Carlo (MCMC) approximation of posterior probability~\cite{plummer_jags_2013}.
	The structure of this code is similar to examples in~\cite{kruschke_doing_2015}.
	
	\paragraph{Check of prediction}
	
	For median values of parameters, we use the mean-square deviation between light pressure force predicted by the corresponding model and light pressure force from ray tracing.
	
	\subsubsection{Bayesian analysis for flat surface}
	
	\paragraph{Data}
	
	In this case, we analyzed two different models of light radiation pressure upon a flat surface.
	The surface is a square area in $O'x'_1x'_2$ plane of the global coordinate system.
	The length of each side is equal to 1m, all sides are parallel to the corresponding axes of the coordinate system, $x'_1,\ x'_2\in[-0.5,0.5],\ x'_3=0$.
	Optical parameters of this surface are uniform and can be specular, diffuse or specular-diffuse with corresponding specularity coefficient:
	$\rho_0=1$ -- reflectivity of the surface, $s_0\in\{0,0.5,1\}$ -- specularity of the surface, $B_0=2/3$ -- Lambertian coefficient of the front side, 3 combinations total.
	We calculated the light radiation pressure vector for all combinations of orientation angles of a light source from these limits:
	\begin{itemize}
		\item $\beta\in\left\{ 9\degree,18\degree,\dots,72\degree,81\degree \right\}$;
		\item $\theta\in\left\{ 0\degree,18\degree,\dots,342\degree,360\degree \right\}$,
	\end{itemize}
	189 combinations total.
	
	The number of rays was 100000 in each simulation.

	\paragraph{Posterior distribution}
	Tab.~\ref{tab:BF} represent the summary of results of MCMC approximation of posterior distribution for different cases. In this table, the Bayesian Factor ($BF$) is the relative posterior probability of a number of a model which is most probable for given data.
	We should note that we used median values instead of the mode of distribution because the speed of convergence for the mode is much lower a than for median.
	
	We performed the analysis for different prior probabilities of a models, e.g. for the specular case we set $P(m=1)=0.01$ and $P(m=2)=0.99$, the MCMC calculated value was $BF'$, and, considering that both prior probabilities should equal to 0.5, according to Bayes rule, the final Bayesian Factor was \cite{kruschke_doing_2015}:
	\[
	BF = BF' \frac{P(m=2)}{P(m=1)}.
	\]
	
	\begin{longtable}{l l l l}
		\caption{Summary of the posterior distributions for different simulations. $BF$ -- \textit{Bayesian Factor} -- posterior probability of the isotropic model divided by posterior probability of the orthotropic model. \textit{Parameters} -- median values of parameters for the most probable model with corresponding 95\% Highest Density Interval (HDI) limits. \textit{Deviation} -- calculated mean-square deviation between Monte Carlo simulated data and calculated data based on median predicted values of the parameters.}\label{tab:BF}\\
		\hline
		\textbf{Case} & \textbf{$BF$} & \textbf{Parameters} & \textbf{Deviation}\\
		\hline
		\endhead
		\hline
		Diffuse plate, & $\approx$0 & $\rho_1=0.443^{+0.148}_{-0.097}$ & \\
		$\rho_0=1,\ s_0=0,\ B_0=2/3$ &  & $\rho_2=0.521^{+0.161}_{-0.130}$ &  \\
		& & $s_1=0.701^{+0.299}_{-0.201}$ & \\
		& & $s_2=0.598^{+0.360}_{-0.209}$ & \\
		& & $k=0.818^{+0.099}_{-0.073}$ & \\
		& & $\sigma_F=0.107^{+0.011}_{-0.009}$ & 0.047\\\hline
		Specular plate, & 297 & $\rho=0.999^{+0.001}_{-0.001}$ & \\
		$\rho_0=1,\ s_0=1$ &  & $s=0.999^{+0.001}_{-0.003}$ &  \\
		& & $B=0.663^{+0.336}_{-0.577}$ & \\
		& & $\sigma_F=0.0148^{+0.0015}_{-0.0015}$ & 0.0108\\
		& & (e.g. Fig. \ref{post:specular_iso}) & \\\hline
		Specular-Diffuse plate, & 99 & $\rho=0.909^{+0.090}_{-0.085}$ & \\
		$\rho_0=1,\ s_0=0.5,\ B_0=2/3$ & & $s=0.511^{+0.062}_{-0.058}$ & \\
		& & $B=0.790^{+0.203}_{-0.131}$ & \\
		& & $\sigma_F=0.0742^{+0.0082}_{-0.0072}$ & 0.0379\\
		\hline
	\end{longtable}
	
	For the diffuse case, the MCMC approximation of posterior distribution showed that the optically orthotropic model of optical parameters is more probable than the isotropic, however, the median values for optical parameters represent the isotropic model very close.
	This problem arises when one tries to compare two models, one of which can represent another~\cite{kruschke_doing_2015}.
	In the subsequent analysis, we will not compare these models using the Bayesian Factor.
	Instead, we will compare these two models by calculating the mean-square deviation between actual data from direct Monte Carlo simulation of ray tracing and predicted data, which will be calculated using isotropic or orthotropic models and the corresponding median values of optical parameters.
	
	\begin{figure}
		\subfloat[Reflectivity]{\includegraphics[width = .5\textwidth]{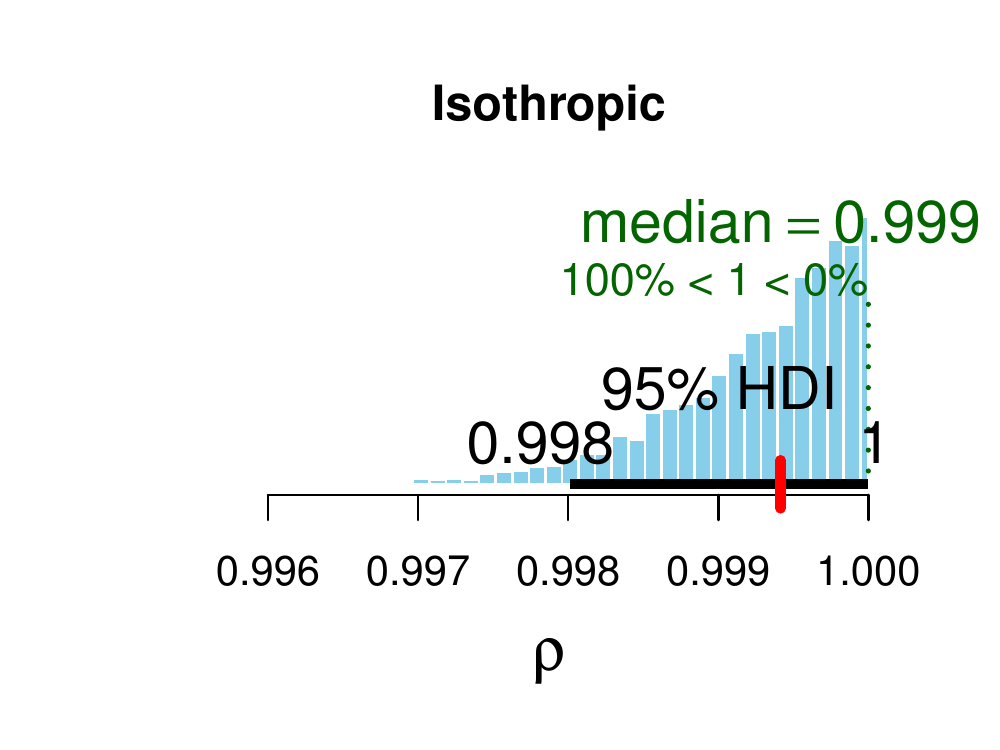}}\    \subfloat[Specularity]{\includegraphics[width = .5\textwidth]{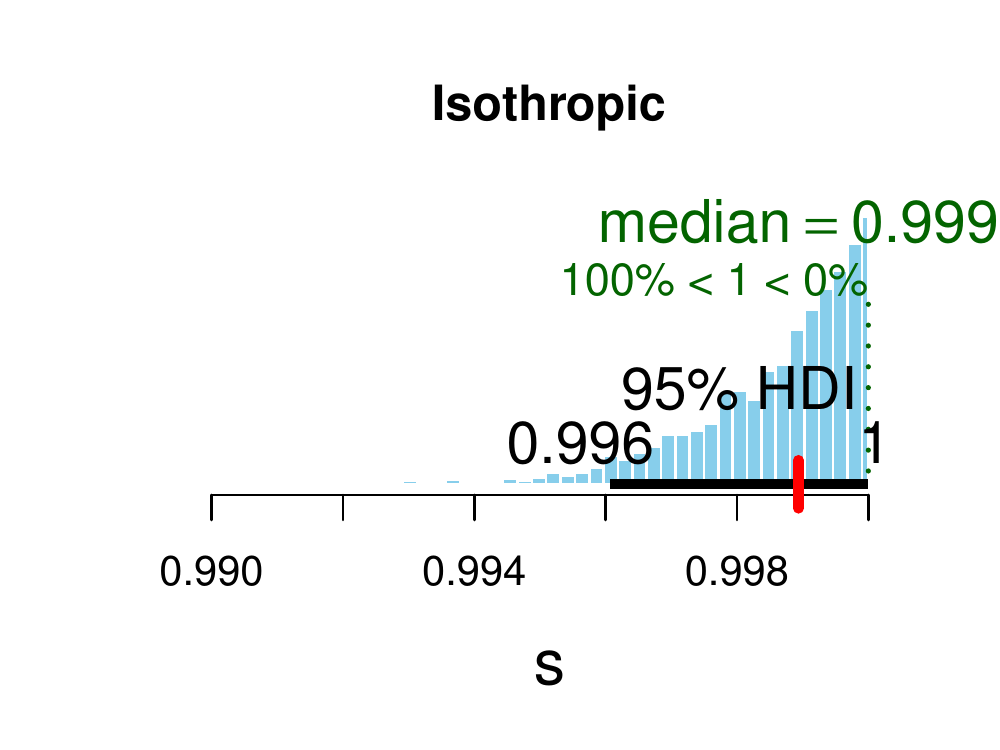}}\
		\subfloat[Lambertian coefficient]{\includegraphics[width = .5\textwidth]{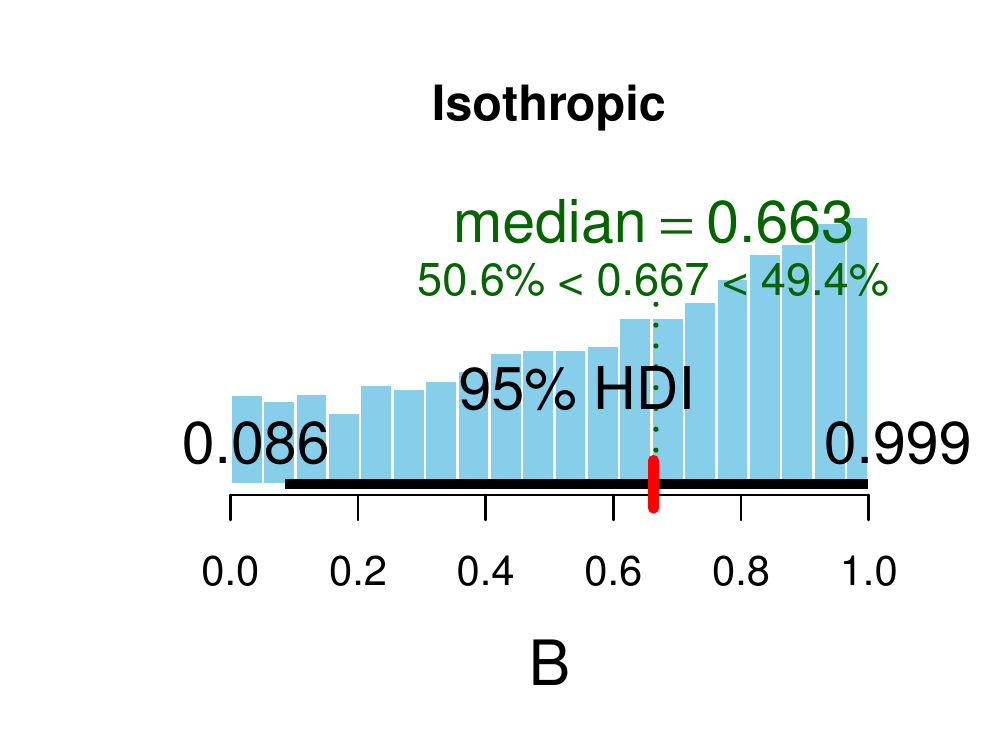}}\
		\subfloat[Standard deviation]{\includegraphics[width = .5\textwidth]{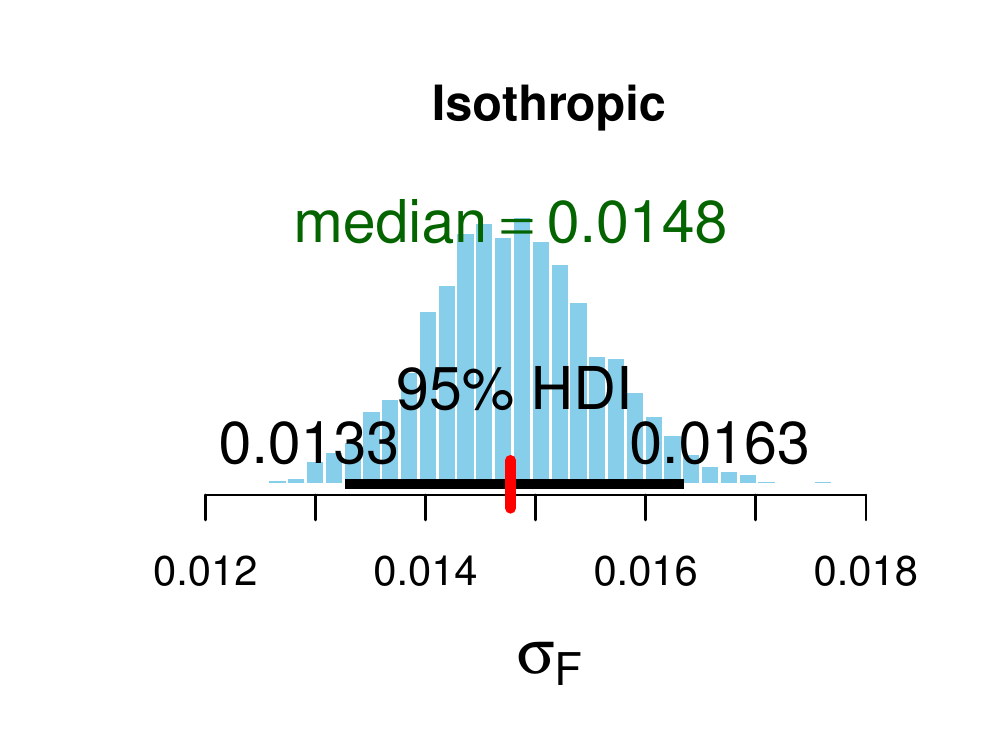}}
		\caption{Markov Chain Monte-Carlo approximation of the posterior distribution of parameters of the model for a flat specular solar sail considering it as an isotropic surface. The vertical scale is a relative probability density function. The wide black horizontal bar represents the highest density interval (HDI) which contains a 95\% of the probability mass. the vertical green dotted line (if any) represents exact values of the parameters, interval shows the ratio of the probability mass to the left of the green line relative to the right. Green labels stand for median values of the approximated density function}
		\label{post:specular_iso}
	\end{figure}

	\subsubsection{Bayesian analysis for curved surface}
	
	\paragraph{Data}
	
	In this case, we analyzed two different models of light radiation pressure upon a surface with waves (wrinkles).
	As it was shown by different authors, both analytically~\cite{epstein_differential_2003,wong_wrinkled_2006-1}, numerically~\cite{wong_wrinkled_2006-2,wang_new_2009,xiao_wrinkle_2011} and experimentally~\cite{wong_wrinkled_2006}, the surface of wrinkled thin membrane can be represented according to the sine or cosine laws.
	
	The magnitude of wrinkles can be calculated by the following formula~\cite{epstein_differential_2003}:
	\[
	x'_3=\delta\sin\frac{2\pi x'_2}{\lambda},
	\]
	Where: $\delta=\delta(x'_1,x'_2)$ -- smooth field of amplitude of wrinkles over the wrinkled domain $\mathcal{D}$; $\lambda=\lambda(x'_1,x'_2)$ -- smooth field of wavelength of wrinkles over the wrinkled domain $\mathcal{D}$.
	
\begin{figure}
		\centering
		\includegraphics[width = 90mm]{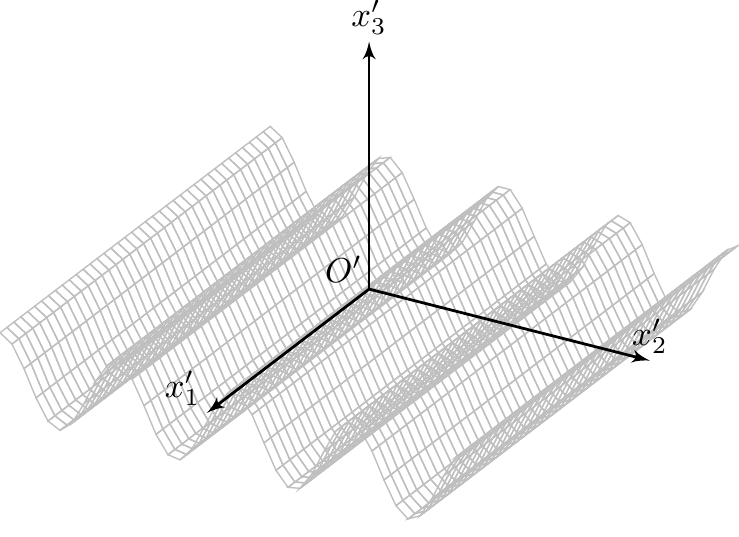}
		\caption{Surface with waves (wrinkles)}\label{f:waves}
	\end{figure}

	We used the following parameters of wrinkled surface: $x'_1,\ x'_2\in[-0.5,0.5]$, $x'_3=\delta(\cos8\pi x'_2 - 1)$.
	There are 4 waves, wavelength $\lambda=const=0.25m$, amplitude $\delta=const=0.2m$, the global frame is shifted along $+x'_3$ by $\delta/2$.
	This surface is shown in Fig.~\ref{f:waves}.
	Optical parameters of this surface are uniform and can be specular, diffuse or specular-diffuse with corresponding specularity coefficient:
	$\rho_0\in\{0.5,1\}$ -- reflectivity of surface, $s_0\in\{0,0.5,1\}$ -- specularity of surface, $B_0=2/3$ -- Lambertian coefficient of the front side, 6 combinations total.
	We calculated the light radiation pressure vector for all combinations of orientation angles of a light source from these limits:
	\begin{itemize}
		\item $\beta\in\left\{ 9\degree,18\degree,\dots,72\degree,81\degree \right\}$;
		\item $\theta\in\left\{ 0\degree,18\degree,\dots,342\degree,360\degree \right\}$,
	\end{itemize}
	189 combinations total.
	
	The number of rays was 100000 in each simulation.
	The ray tracing results are available in the Mendeley Data~\cite{nerovny_data_2017}.
	
	\paragraph{Posterior distribution}
	
	Tab.~\ref{tab:cosine} represents the summary of the MCMC approximation of parameters of models for different values of $\rho_0$ and $s_0$.
	There are several cases for which the mean-square deviation of an orthotropic surface is less than for an isotropic surface; i.e., the orthotropic model is more accurate.
	This fact confirms that the model of an optically orthotropic surface proposed in this paper can be suitable for practical applications, providing a more accurate representation than an isotropic model, at least if not considering thermal emission.
	
	\begin{longtable}{l l l}
		\caption{Summary of posterior distributions for different simulations for surface with waves with different optical parameters (\textit{case} column). \textit{Isotropic} -- summary of MCMC approximation considering an isotropic model of optical parameters, value is a median, limits are 95\% HDI. \textit{Orthotropic} -- summary of MCMC approximation considering an orthotropic model of optical parameters, value is a median, limits are 95\% HDI.  \textit{Deviation} -- calculated mean-square deviation between Monte Carlo simulated data and calculated data based on median predicted values of the parameters.}\label{tab:cosine}\\
		\hline
		\textbf{Case} & \textbf{Isotropic} & \textbf{Orthotropic}\\
		\hline
		\endhead
		\hline
		$\rho_0=1,\ s_0=1$ & $\rho=0.870^{+0.130}_{-0.105}$ & $\rho_1=0.749^{+0.233}_{-0.156}$ \\
		& $s=0.516^{+0.144}_{-0.136}$ & $\rho_2=0.729^{+0.219}_{-0.174}$ \\
		& $B=0.770^{+0.230}_{-0.200}$ & $s_1=0.803^{+0.191}_{-0.256}$ \\
		& $\sigma_F=0.281^{+0.030}_{-0.028}$ & $s_2=0.822^{+0.178}_{-0.297}$ \\
		&  & $k=0.186^{+0.103}_{-0.085}$ \\
		&  & $\sigma_F=0.274^{+0.025}_{-0.024}$ \\
		\textbf{Deviation:} & $0.108$ & $0.093$ \\\hline
		$\rho_0=1,\ s_0=0.5,\ B_0=2/3$ & $\rho=0.574^{+0.325}_{-0.146}$ & $\rho_1=0.291^{+0.323}_{-0.195}$ \\
		& $s=0.010^{-0.034}_{-0.010}$ & $\rho_2=0.476^{+0.324}_{-0.191}$ \\
		& $B=0.776^{+0.224}_{-0.281}$ & $s_1=0.179^{+0.452}_{-0.179}$ \\
		& $\sigma_F=0.163^{+0.017}_{-0.017}$ & $s_2=0.222^{+0.394}_{-0.154}$ \\
		&  & $k=1.870^{+3.850}_{-0.850}$ \\
		&  & $\sigma_F=0.122^{+0.016}_{-0.011}$ \\
		\textbf{Deviation:} & $0.091$ & $0.047$ \\\hline
		$\rho_0=1,\ s_0=0,\ B_0=2/3$ & $\rho=0.611^{+0.327}_{-0.196}$ & $\rho_1=0.104^{+0.034}_{-0.023}$ \\
		& $s=0.005^{+0.018}_{-0.005}$ & $\rho_2=0.369^{+0.083}_{-0.085}$ \\
		& $B=0.707^{+0.293}_{-0.240}$ & $s_1=0.677^{+0.308}_{-0.240}$ \\
		& $\sigma_F=0.190^{+0.021}_{-0.018}$ & $s_2=0.331^{+0.197}_{-0.136}$ \\
		&  & $k=2.440^{+0.790}_{-0.510}$ \\
		&  & $\sigma_F=0.097^{+0.009}_{-0.009}$ \\
		(e.g. Fig.~\ref{post:compare}) & & \\
		\textbf{Deviation:} & $0.101$ & $0.039$ \\\hline
		$\rho_0=0.5,\ s_0=1$ & $\rho=0.314^{+0.488}_{-0.169}$ & $\rho_1=0.047^{+0.103}_{-0.047}$ \\
		& $s=0.063^{+0.160}_{-0.063}$ & $\rho_2=0.211^{+0.082}_{-0.123}$ \\
		& $B=0.479^{+0.453}_{-0.337}$ & $s_1=0.094^{+0.462}_{-0.94}$ \\
		& $\sigma_F=0.139^{+0.015}_{-0.013}$ & $s_2=0.296^{+0.406}_{-0.207}$ \\
		&  & $k=2.330^{+1.680}_{-1.130}$ \\
		&  & $\sigma_F=0.123^{+0.013}_{-0.012}$ \\
		\textbf{Deviation:} & $0.068$ & $0.040$ \\\hline
		$\rho_0=0.5,\ s_0=0.5,\ B_0=2/3$ & $\rho=0.332^{+0.505}_{-0.175}$ & $\rho_1=0.057^{+0.077}_{-0.057}$ \\
		& $s=0.008^{+0.039}_{-0.008}$ & $\rho_2=0.074^{+0.133}_{-0.031}$ \\
		& $B=0.519^{+0.456}_{-0.334}$ & $s_1=0.218^{+0.539}_{-0.206}$ \\
		& $\sigma_F=0.120^{+0.013}_{-0.012}$ & $s_2=0.448^{+0.456}_{-0.346}$ \\
		& & $k=4.450^{+2.130}_{-1.120}$ \\
		& & $\sigma_F=0.081^{+0.007}_{-0.008}$ \\
		\textbf{Deviation:} & $0.074$ & $0.037$ \\\hline
		$\rho_0=0.5,\ s_0=0,\ B_0=2/3$ & $\rho=0.400^{+0.482}_{-0.192}$ & $\rho_1=0.041^{+0.025}_{-0.013}$ \\
		& $s=0.006^{+0.024}_{-0.006}$ & $\rho_2=0.121^{+0.038}_{-0.036}$ \\
		& $B=0.580^{+0.387}_{-0.346}$ & $s_1=0.806^{+0.194}_{-0.264}$ \\
		& $\sigma_F=0.137^{+0.015}_{-0.013}$ & $s_2=0.524^{+0.146}_{-0.206}$ \\
		& & $k=3.480^{+1.230}_{-1.170}$ \\
		& & $\sigma_F=0.073^{+0.008}_{-0.006}$ \\
		\textbf{Deviation:} & $0.083$ & $0.031$ \\\hline
	\end{longtable}
	
	\begin{figure}
		\subfloat[$\beta=9\degree$]{\includegraphics[width = .32\textwidth]{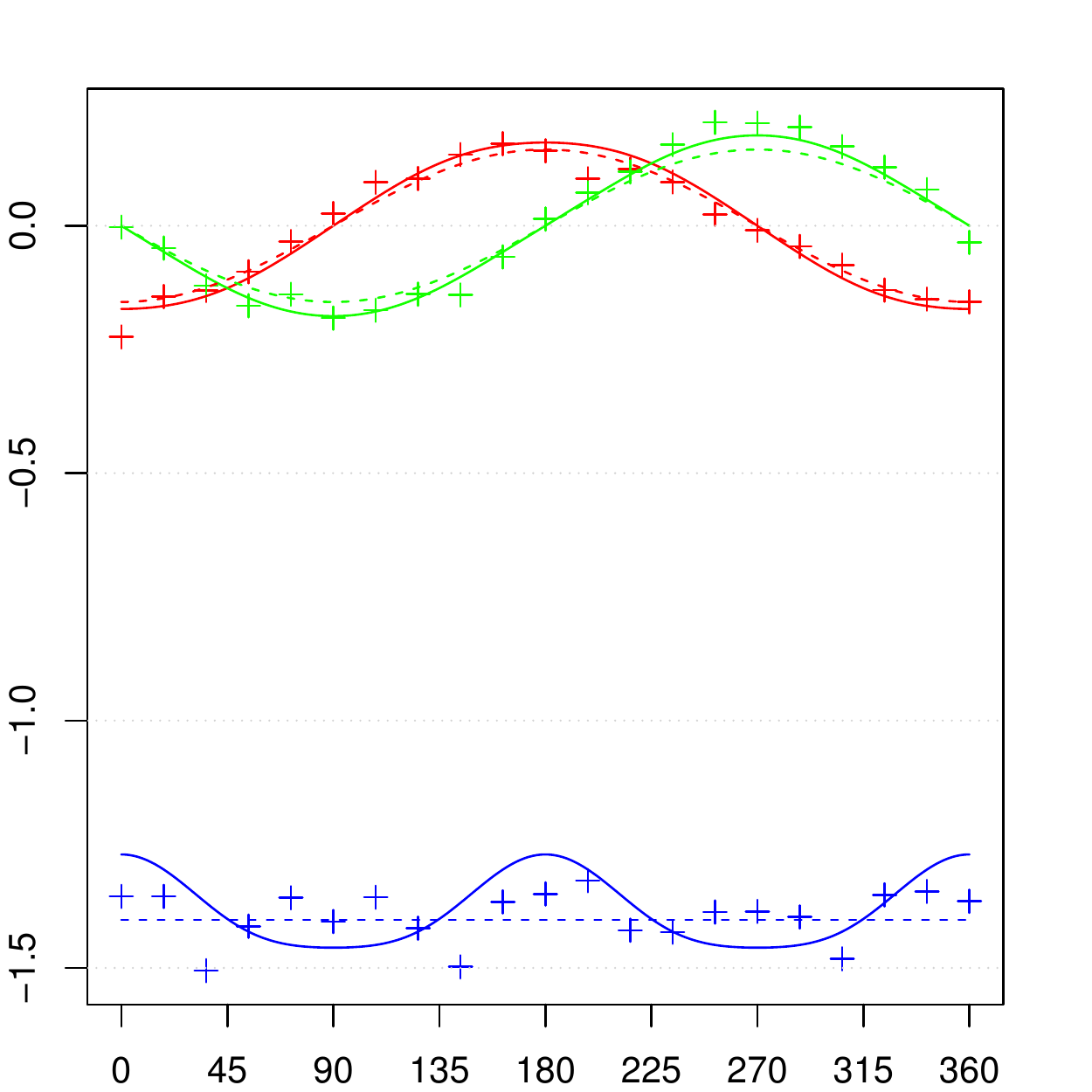}}\    \subfloat[$\beta=18\degree$]{\includegraphics[width = .32\textwidth]{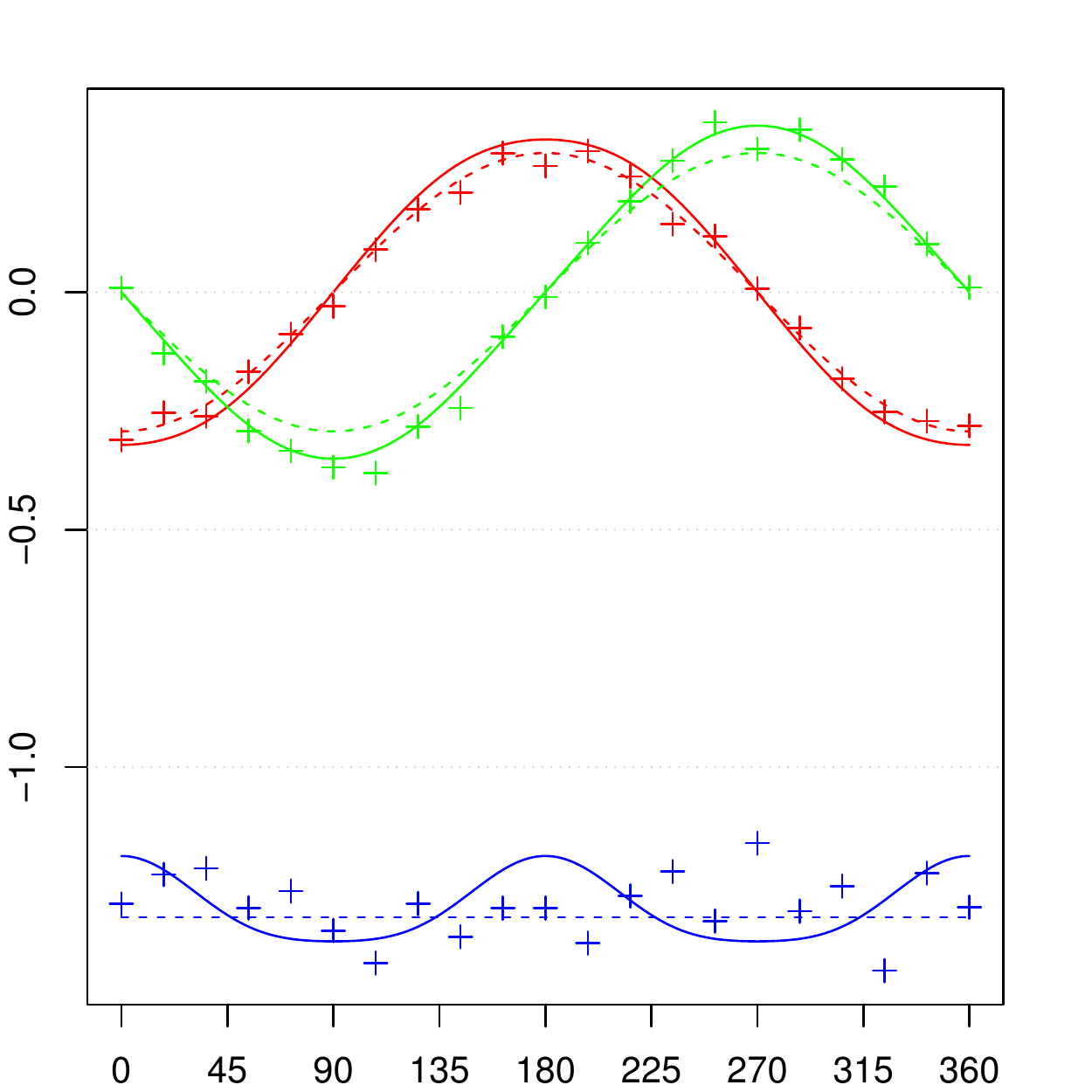}}\
		\subfloat[$\beta=27\degree$]{\includegraphics[width = .32\textwidth]{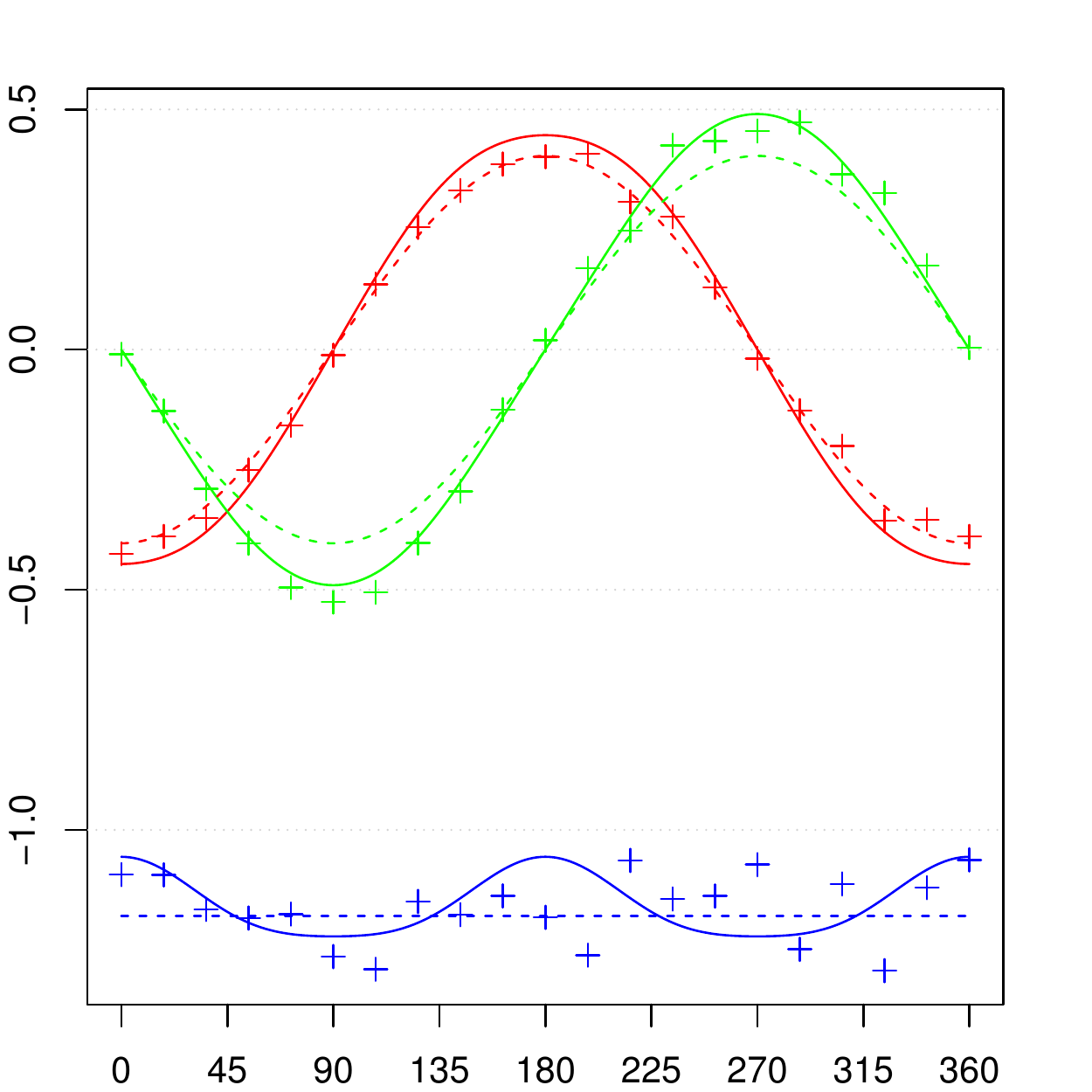}}\
		\subfloat[$\beta=36\degree$]{\includegraphics[width = .32\textwidth]{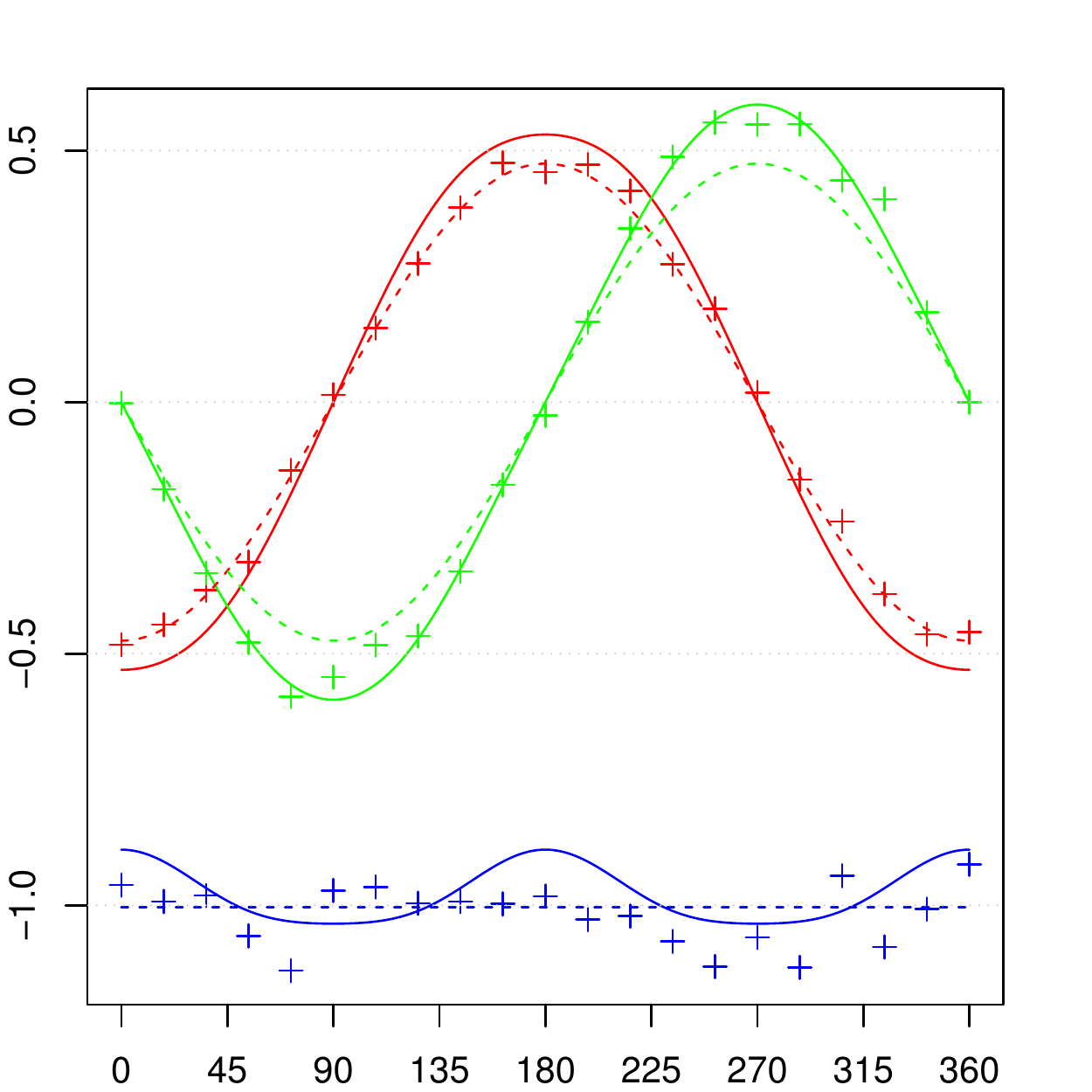}}\
		\subfloat[$\beta=45\degree$]{\includegraphics[width = .32\textwidth]{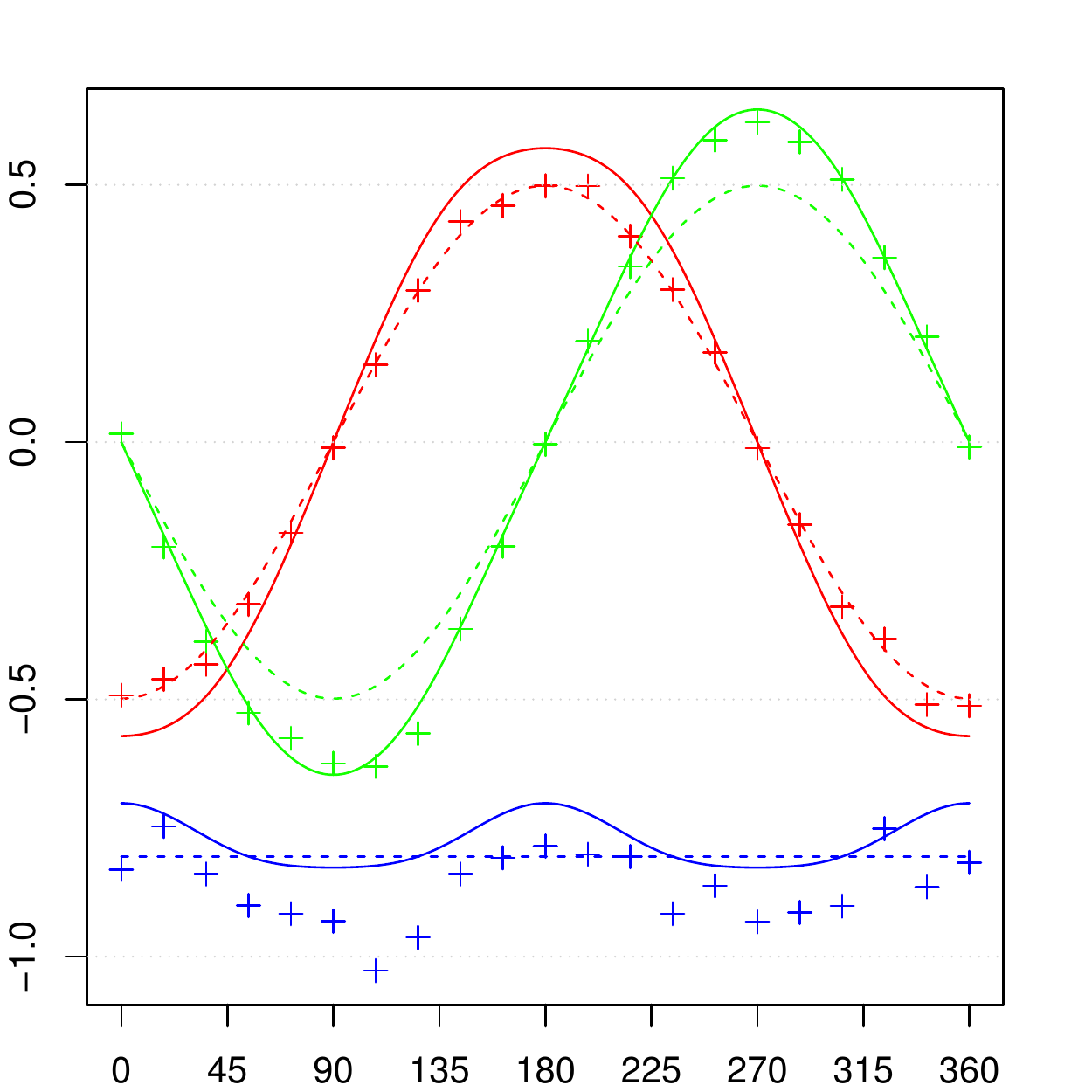}}\    \subfloat[$\beta=54\degree$]{\includegraphics[width = .32\textwidth]{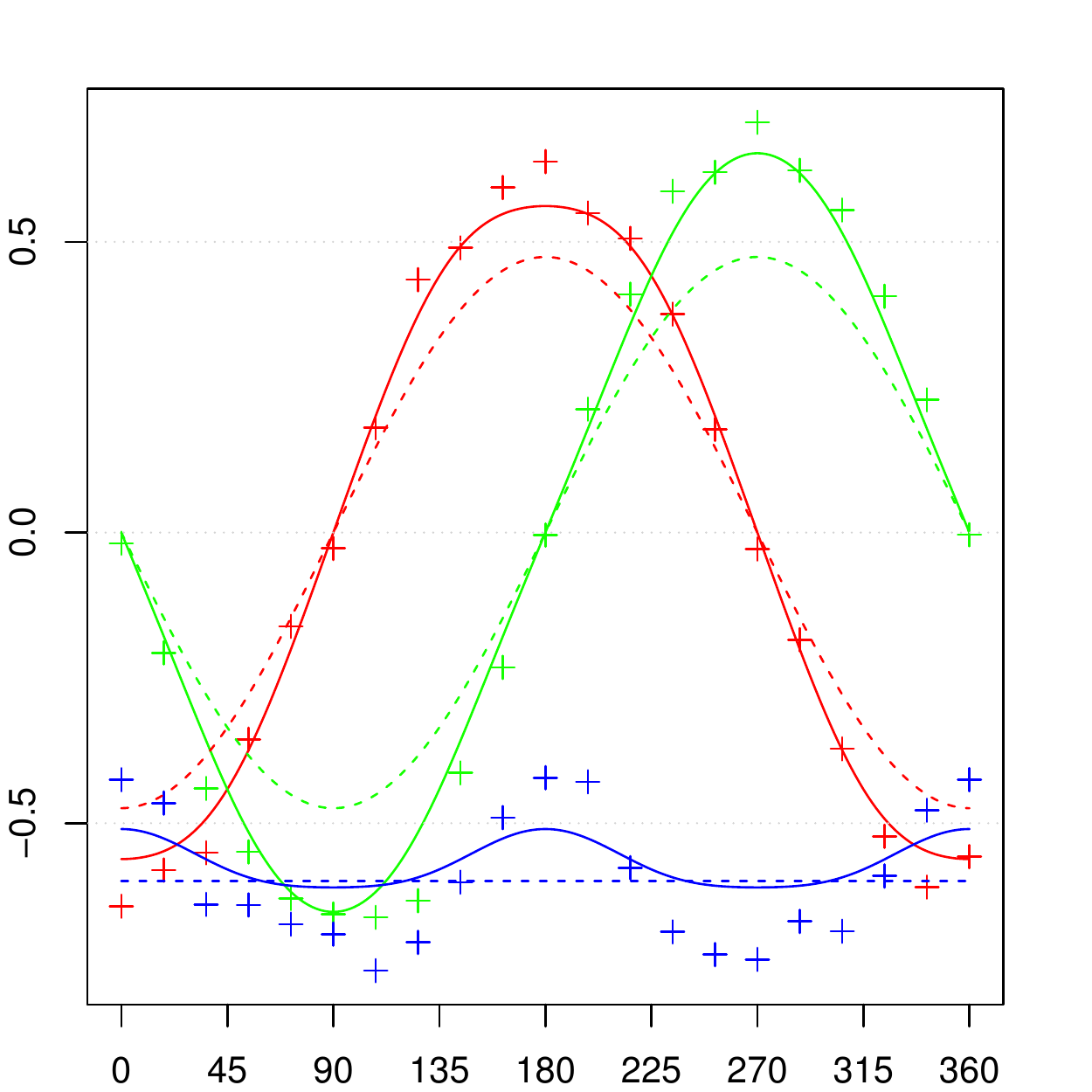}}\
		\subfloat[$\beta=63\degree$]{\includegraphics[width = .32\textwidth]{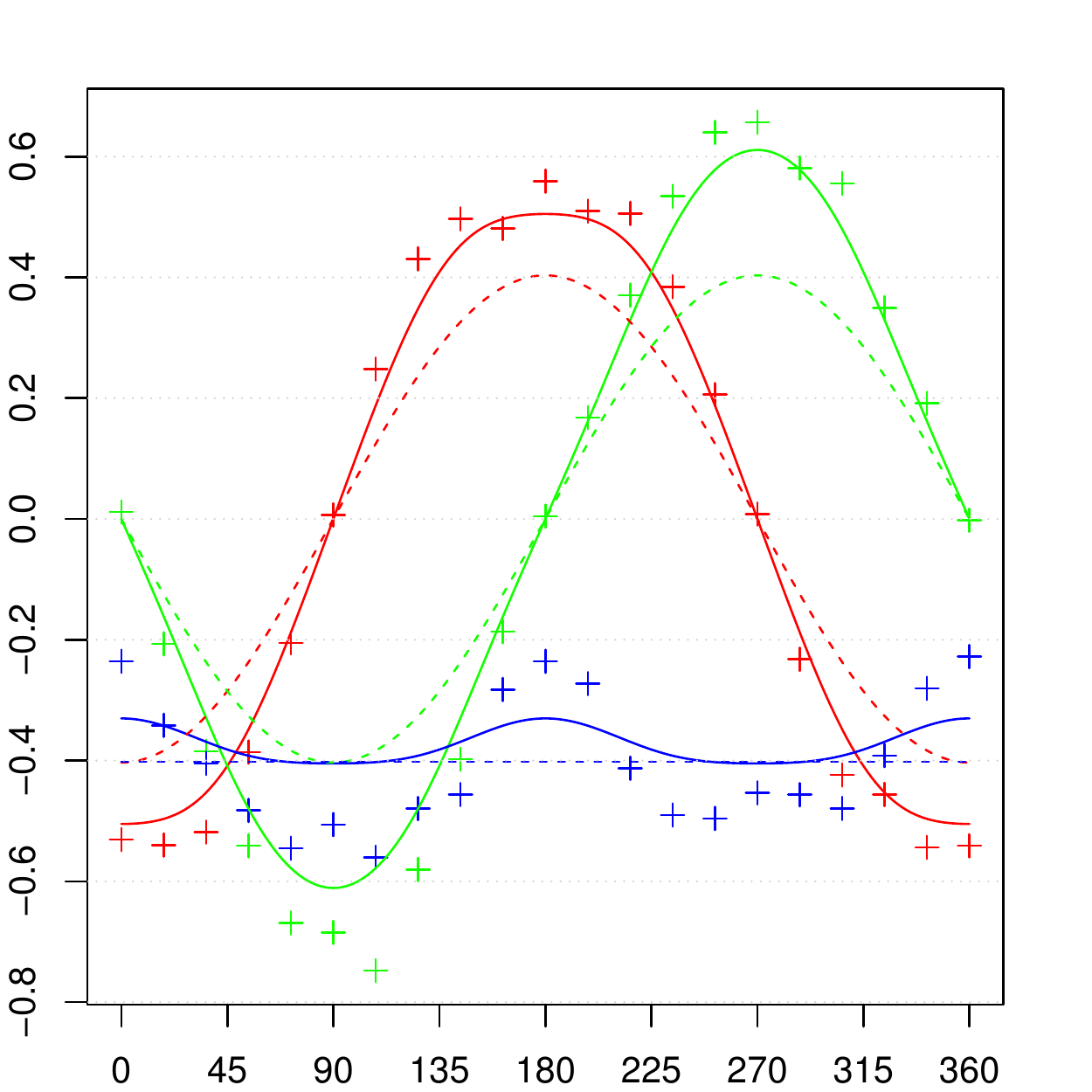}}\
		\subfloat[$\beta=72\degree$]{\includegraphics[width = .32\textwidth]{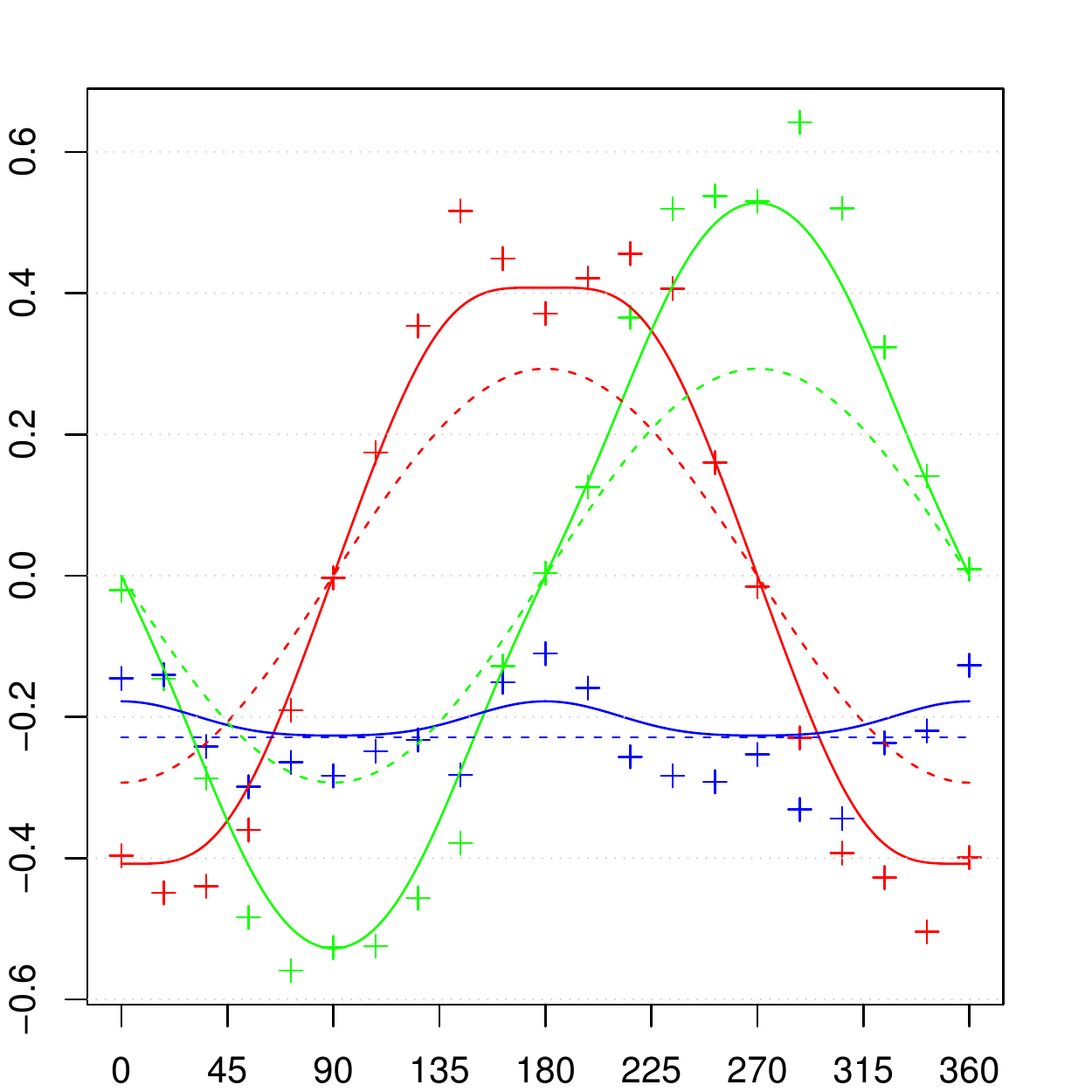}}\
		\subfloat[$\beta=81\degree$]{\includegraphics[width = .32\textwidth]{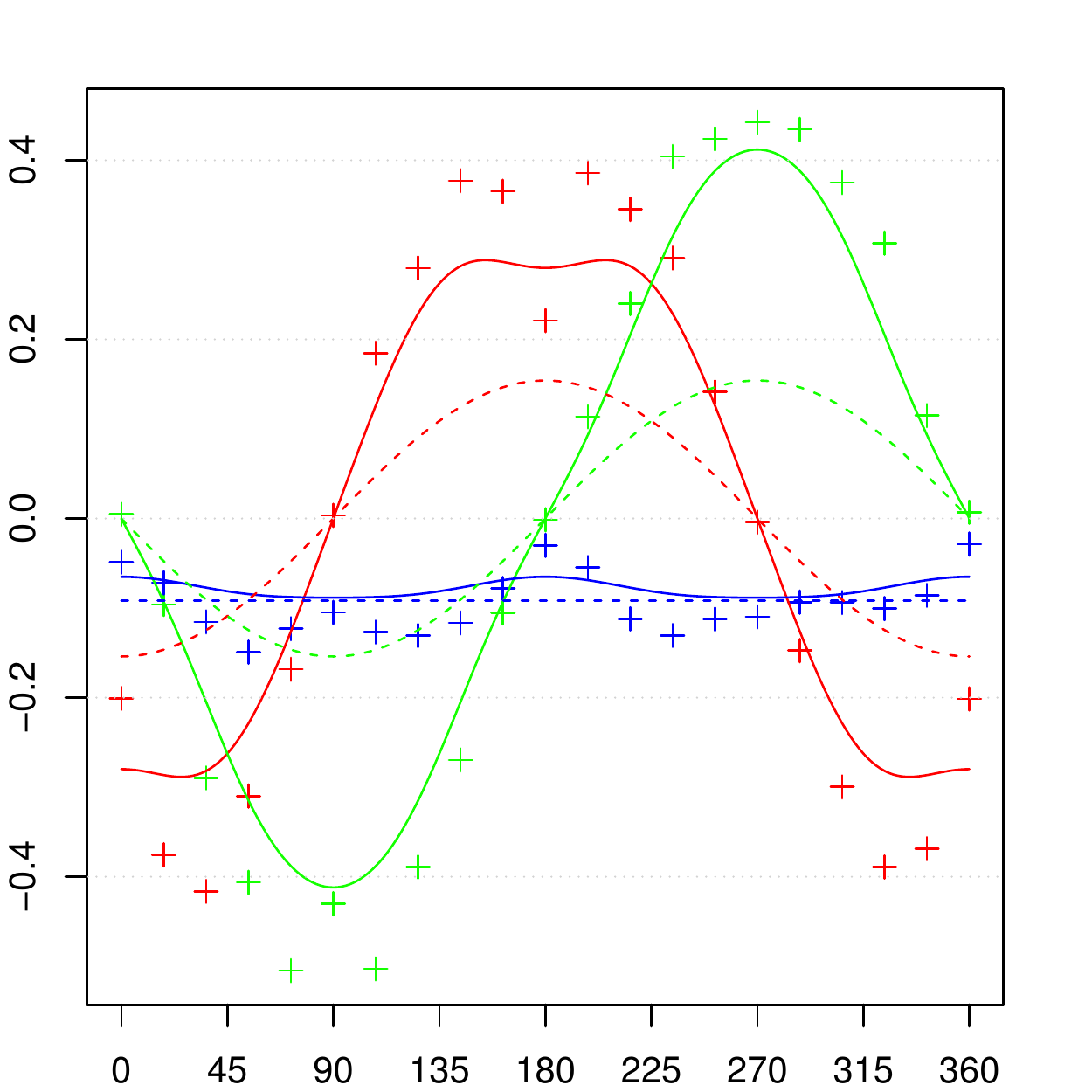}}
		\caption{Comparison between ray tracing results, isotropic model and orthotropic model for surface with waves, $\rho=1,\ s=0,\ B=2/3$. Vertical axis -- light pressure force, $m^2$. Horizontal axis -- $\theta$. Red lines and dots -- $F'_1,\ m^2$ (projection on $O'x'_1$ of global frame). Green lines and dots -- $F'_2,\ m^2$ (projection on $O'x'_2$ of global frame). Blue lines and dots -- $F'_3,\ m^2$ (projection on $O'x'_3$ of global frame). Dots -- ray tracing results. Solid lines -- orthotropic model based on median values of parameters. Dashed lines -- isotropic model based on median values of parameters}
		\label{post:compare}
	\end{figure}
	
	\subsection{Conformity with Generalized Sail Model}
	
	The other difficulty with a model of an optically orthotropic surface is the fact that it is not well fitted with the Generalized Sail Model~\cite{rios_reyes_generalized_2005}, since it uses an additional vector of an orientation of optical axes $\unit{m}$, as soon as there is no such vector in the original GSM.
	One of the possible ways of dealing with this may be the expansion of terms with $\unit{m}$ into some power series.
	Further analytical separation of $\unit{m}$ from $\unit{s}$ may be accomplished in the same way as it was done for $\unit{n}$ and $\unit{s}$ in~\cite{nerovny_representation_2017}.
	Thus conformity with GSM needs to be investigated.
	
	For practical applications without the GSM, it is possible to utilize the orthotropic model for better accuracy after derivation of model parameters for given surface.
	
	\section{Acknowledgements}
	
	The authors would like to thank assistant Dmitry A. Goncharov from the department ``Theoretical Mechanics'' of BMSTU and Evgeny S. Golubev from Astro Space Center of the P.N. Lebedev Physical Institute of the Russian Academy of Sciences for their valuable advice and discussions.
	The authors also would like to thank Mark A. Bowman and Eugene Chebezov from the Flight Dynamics Division, NASA Johnson Space Center, for their help in proofreading of the article.
	
	Contributions: N.A.~Nerovny -- model and analysis, development of srp2 software, I.E.~Lapina -- Bayesian analysis, A.S.~Grigorjev -- srp2 software.
	
	This work was done during the development of BMSTU-Sail Space Experiment~\cite{bmstu_iac_2011,bmstu_knts}.
	
	\bibliographystyle{apa} 	
	\bibliography{nerovny}
	
\end{document}